\documentclass[twocolumn,draft,prb]{revtex4}
\usepackage{graphicx}% Include figure files
\usepackage{dcolumn}% Align table columns on decimal point
\usepackage{bm}% bold math
\usepackage{subfigure}
\usepackage[USenglish]{babel}
\usepackage{amssymb,amsmath}
\usepackage{mathtools}
\usepackage{relsize}
\usepackage{lmodern}
\usepackage{slantsc}
\usepackage{scalefnt}
\usepackage{tikz}
\usetikzlibrary{arrows,calc,scopes}
\usetikzlibrary{shapes.geometric}
\usepackage{color}
\allowdisplaybreaks[1]

\newcommand{\be}{\begin{equation}}
\newcommand{\ee}{\end{equation}}
\newcommand{\bse}{\begin{subequations}}
\newcommand{\ese}{\end{subequations}}
\newcommand{\bpm}{\begin{pmatrix}}
\newcommand{\epm}{\end{pmatrix}}
\newcommand{\bmm}{\begin{matrix}}
\newcommand{\emm}{\end{matrix}}
\newcommand{\Z}{\mathbb{Z}}
\newcommand{\N}{\mathcal{N}}
\newcommand{\F}{\mathfrak{F}}
\newcommand{\ii}{\mathrm{i}}
\newcommand{\e}{\mathrm{e}}
\newcommand{\s}[2][]{_{\textsl{\textsc{#2}}{#1}}}
\makeatletter
\newcommand*{\Relbarfill@}{\arrowfill@\Relbar\Relbar\Relbar}
\newcommand*{\xeq}[2][]{\ext@arrow 0055\Relbarfill@{#1}{#2}}
\makeatother
\newtheorem{theorem}{Theorem}

\newcommand{\BLvert}{\Biggl\vert\bmm\vspace{-3pt}\scalefont{0.6}} %the big \vert on the right
\newcommand{\Brangle}{\emm\Biggr\rangle}

\newcommand{\sline}{\begin{tikzpicture}[scale=0.6,baseline]
\draw (0,0) -- (0.5,0) node[above,inner sep=0, outer sep=1] {\scalebox{0.8}{$i$}} -- (1,0);
\draw[->,>=stealth',line width=0.01pt] (0.4,0) -- (0.5,0);
\end{tikzpicture}}%draw a straight line with label i

\newcommand{\curveline}{\begin{tikzpicture}[scale=0.8,baseline]
\draw (0,0) -- ++(0.25,0) -- ++(300:0.25) -- ++(0.25,0) -- ++(60:0.25) -- +(0.25,0);
\node[inner sep=0,outer sep=1] at(0.5,0) {\scalebox{0.8}{$i$}};
\draw[->,>=stealth',line width=0.01pt] (0,0) ++(0.25,0) ++(300:0.25) -- ++(0.15,0);
\end{tikzpicture}}%draw a curved line with label i

\newcommand{\Ygraph}[4][1]{
\begin{tikzpicture}[scale=0.6,baseline]
  \draw [->,>=stealth',line width=0.01pt] (30:0.1) -- (0,0) ; %draw an arrow thin enough
    \draw (30:1) -- (0,0) ; %draw the line
  \draw [->,>=stealth',line width=0.01pt] (150:0.1) -- (0,0); %draw an arrow thin enough
    \draw (150:1) -- (0,0); %draw the line
  \node at(-0.5,0.01) {$#2$};
  \node at(0.5,0.01) {$#4$};
  \ifnum #1=1 {
   \draw [->,>=stealth',line width=0.01pt] (0,-1/20) -- (0,0); %draw an arrow thin enough
   \node at(-0.3,-0.5) {$#3$};
  }
  \else{
    \draw [<-,>=stealth',line width=0.01pt] (0,-1/2) -- (0,0); %draw an arrow thin enough
    \node at(-0.4,-0.5) {$#3$};
    }
  \fi
    \draw (0,-1) -- (0,0); %draw the line
  \end{tikzpicture}
  }

\newcommand{\Hgraph}[3][1]{
  \begin{tikzpicture}[scale=0.6]
    \coordinate (c) at (0,0);
    \coordinate (l) at (-0.7, 0);
    \coordinate (r) at ($ (c) ! -1 ! (l) $);
    \coordinate (ul) at (-0.95,0.75);
    \coordinate (lr) at ($ (c) ! -1 ! (ul) $);
    \coordinate (ll) at (-0.95,-0.75);
    \coordinate (ur) at ($ (c) ! -1 ! (ll) $);
    \draw (l) -- (r) ; %draw the horizontal line
    \draw (ul) -- (l); %draw the upper left line
    \draw (ll) -- (l); %draw the lower left line
    \draw (lr) -- (r); %draw the lower right line
    \draw (ur) -- (r); %draw the upper right line
    \ifnum #1=1 {
    \node[right] at($ (ul) ! .3 ! (l) $) {$#2_{\text{\scalebox{0.7}{$1$}}} $};
    \node[right] at($ (ll) ! .2 ! (l) $) {$#2_{\text{\scalebox{0.7}{$2$}}} $};
    \node[left] at($ (lr) ! .2 ! (r) $) {$#2_{\text{\scalebox{0.7}{$3$}}} $};
    \node[left] at($ (ur) ! .3 ! (r) $) {$#2_{\text{\scalebox{0.7}{$4$}}} $};
    \node[below] at($ (c) ! .15 ! (0,0.5) $) {$#3 $};}
    \fi
    \draw[<-,>=stealth', line width=0.01pt] (l) -- (r) ;
    \draw[->,>=stealth', line width=0.01pt] (ul) -- (l);
    \draw[->,>=stealth', line width=0.01pt] (ll) -- (l);
    \draw[->,>=stealth', line width=0.01pt] (lr) -- (r);
    \draw[->,>=stealth', line width=0.01pt] (ur) -- (r);
   \end{tikzpicture}
  }

 \newcommand{\Xgraph}[3][1]{
  \begin{tikzpicture}[scale=0.6]
    \coordinate (c) at (0,0);
    \coordinate (u) at (0, 0.6);
    \coordinate (d) at ($ (c) ! -1 ! (u) $);
    \coordinate (ul) at (-0.85,0.85);
    \coordinate (lr) at ($ (c) ! -1 ! (ul) $);
    \coordinate (ll) at (-0.85,-0.85);
    \coordinate (ur) at ($ (c) ! -1 ! (ll) $);
    \draw (u) -- (d) ; %draw the horizontal line
    \draw (ul) -- (u); %draw the upper left line
    \draw (ll) -- (d); %draw the lower left line
    \draw (lr) -- (d); %draw the lower right line
    \draw (ur) -- (u); %draw the upper right line
    \ifnum #1=1 {
    \node[below] at($ (ul) ! .3 ! (u) $) {$#2_{\text{\scalebox{0.7}{$1$}}} $};
    \node[above] at($ (ll) ! .2 ! (d) $) {$#2_{\text{\scalebox{0.7}{$2$}}} $};
    \node[above] at($ (lr) ! .2 ! (d) $) {$#2_{\text{\scalebox{0.7}{$3$}}} $};
    \node[below] at($ (ur) ! .2 ! (u) $) {$#2_{\text{\scalebox{0.7}{$4$}}} $};
    \node[right] at($ (-0.5,0) ! 0.8 ! (c) $) {$#3 $}; }
    \fi
    \draw[<-,>=stealth', line width=0.01pt] (u) -- (d) ;
    \draw[->,>=stealth', line width=0.01pt] (ul) -- (u);
    \draw[->,>=stealth', line width=0.01pt] (ll) -- (d);
    \draw[->,>=stealth', line width=0.01pt] (lr) -- (d);
    \draw[->,>=stealth', line width=0.01pt] (ur) -- (u);
   \end{tikzpicture}
  }

\newcommand{\Psix}[3][1]{
\begin{tikzpicture}[scale=0.8]
\node[name=s, regular polygon, regular polygon sides=6, minimum size=1cm, outer sep=0pt ,draw] at (0,0) {}; %draw the plaquette
\foreach \anchor/\x/\y /\xx/\yy /\b in
{corner 1/0.17/0.17*1.732/-0.11/0.18/1, corner 2/-0.17/0.17*1.732/0.07/0.18/2, corner 3/-0.34/0/-0.15/-0.18/3, corner 4/-0.17/-0.17*1.732/-0.22/-0.05/4, corner 5/0.17/-0.17*1.732/0.2/-0.05/5, corner 6/0.34/0/0.15/-0.18/6}
{
 \draw[shift=(s.\anchor)] (0,0) -- (\x,\y) node at(\xx,\yy) {$#2_{\text{\scalebox{0.7}{$\b$}}}$};
 \ifnum #1=1
 \draw[shift=(s.\anchor),<-,>=stealth', line width=0.01pt] (s.\anchor) -- (\x,\y);
 \fi
 }
\foreach \anchor/\xx/\yy /\a in
{side 1/0/-0.18/1, side 2/-0.18/0.05/2, side 3/0.15/0.05/3, side 4/0/-0.18/4, side 5/-0.18/0.05/5, side 6/0.15/0.05/6}
 \draw[shift=(s.\anchor)]  node at(\xx,\yy) {$#3_{\text{\scalebox{0.7}{$\a$}}}$};
\ifnum #1=1{
  \foreach \anchorr/\anchorf in
   {corner 1/corner 2, corner 2/corner 3, corner 3/corner 4, corner 4/corner 5, corner 5/corner 6, corner 6/corner 1}
   \draw[shift=(s.\anchorr), ->, >=stealth', line width=0.01pt]  (s.\anchorr) -- (s.\anchorf);}
 \else {
  \foreach \anchorb/\anchorw in
   {corner 1/corner 2, corner 3/corner 4, corner 5/corner 6} {
   \node[fill=black, circle, minimum size=2.5, inner sep=0, outer sep=0, draw] at(s.\anchorb) {};
   \node[fill=white, circle, minimum size=2.5, inner sep=0, outer sep=0, draw] at(s.\anchorw) {};}
}
\fi
\end{tikzpicture}
}
\newcommand{\thetaGraph}{
  \begin{tikzpicture}[scale=0.6]
  \draw (2,0) arc (0:180:1 and 0.8) ;
  \draw [<-,>=stealth', line width=0.01pt] (2,0) arc (0:180:1 and 0.8);
  \draw (0,0) arc (180:360:1 and 0.8);
  \draw [->,>=stealth', line width=0.01pt] (0,0) arc (180:360:1 and 0.8);
  \draw (2,0) -- (0,0);
  \draw [->,>=stealth', line width=0.01pt] (2,0) -- (0,0);
  \node[fill=white, inner sep=0,outer sep=0] at(1,0.8) {$i$};
  \node[fill=white, inner sep=0,outer sep=0] at(1,0) {$j$};
  \node[fill=white, inner sep=0,outer sep=0] at(1,-0.8) {$k$};
  \end{tikzpicture}}

\newcommand{\bubbleGraph}[4]{\begin{tikzpicture}[scale=0.6]
  \draw (1,0) arc (0:180:0.5 and 0.4) ;
  \draw [->,>=stealth', line width=0.01pt] (1,0) arc (0:90:0.5 and 0.4) node[below left] {\scalebox{0.5}{$#3$}};
  \draw (0,0) arc (180:360:0.5 and 0.4);
  \draw [->,>=stealth', line width=0.01pt] (0,0) arc (180:270:0.5 and 0.4) node[above right] {\scalebox{0.5}{$#4$}};
  \draw (-0.5,0) -- (0,0);
  \draw [->,>=stealth', line width=0.01pt] (-0.5,0) -- (-0.2,0) node[above, inner sep=0,outer sep=1] {\scalebox{0.5}{$#1$}};
  \draw (1,0) -- (1.5,0);
  \draw [->,>=stealth', line width=0.01pt] (1,0) -- (1.35,0) node[above, inner sep=0,outer sep=1] {\scalebox{0.5}{$#2$}};
  \end{tikzpicture}}

\newcommand{\loopGraph}[1]{\begin{tikzpicture}[scale=0.4,baseline]
 \draw (1,0) arc (0:360:0.5) node[left] {\scalebox{0.8}{$#1$}};
 \draw[->, >=stealth',line width=0.01pt] (1,0) arc (0:10:0.5); \end{tikzpicture}}

\newcommand{\shadowGraph}{\begin{tikzpicture}[scale=0.5,baseline]
 \draw[fill=gray,rounded corners,ultra thick, gray] (-0.4,0.5) rectangle (0.4,-0.5);
 \end{tikzpicture}}

\newcommand{\honeycomb}[2]{\begin{tikzpicture}[scale=0.5]
   \clip[draw] (0,0) rectangle (2*1.732,-3);
\def\hexagonpath{ +(30:1) -- +(90:1)  -- +(150:1) -- +(210:1) -- +(270:1) -- +(330:1)  -- cycle }
\foreach \x in {0,...,#1}
  \foreach \y in {0,...,#2} {
    \ifodd\x
     \draw (0,0) ++(0,-1/2-3*\x/2) ++(30:1) ++(30:\y * 2) ++(0,-\y) \hexagonpath;
    \else\draw (0,0) ++(0,-3*\x/2) ++(30:\y * 2) ++(0,-\y) \hexagonpath;
    \fi }
\end{tikzpicture}}

\newcommand{\dualityGraph}{\begin{tikzpicture}[scale=1.2]
\node[name=s, regular polygon, regular polygon sides=6, minimum size=40, outer sep=0pt, inner sep=0, draw] at (0,0) {}; %draw the plaquette
\foreach \anchor/\x/\y /\xx/\yy /\b in
{corner 1/0.3/0.3*1.732/0.38/0.38/1, corner 2/-0.3/0.3*1.732/-0.34/0.38/2, corner 3/-0.6/0/-0.48/0.08/3, corner 4/-0.3/-0.3*1.732/-0.36/-0.38/4, corner 5/0.3/-0.3*1.732/0.36/-0.38/5, corner 6/0.6/0/0.5/0.08/6}
 \draw[shift=(s.\anchor)] (0,0) -- (\x,\y) node at(\xx,\yy) {$l_{\text{\scalebox{0.7}{$\b$}}}$};
\foreach \anchor/\xx/\yy /\a in
{side 1/0.15/-0.14/1, side 2/-0.06/0.16/2, side 3/0.05/0.16/3, side 4/-0.1/-0.13/4, side 5/-0.17/-0.04/5, side 6/0.17/-0.07/6}
 \draw[shift=(s.\anchor)]  node at(\xx,\yy) {$e_{\text{\scalebox{0.7}{$\a$}}}$};

\foreach \anchorb/\anchorw in {corner 1/corner 2, corner 3/corner 4, corner 5/corner 6}
   {\node[fill=black, circle, minimum size=2.5, inner sep=0, outer sep=0, draw] at(s.\anchorb) {};
   \node[fill=white, circle, minimum size=2.5, inner sep=0, outer sep=0, draw] at(s.\anchorw) {};}

\node[name=a, fill=red, circle, double, minimum size=2, inner sep=0, outer sep=0, draw=red] at (0,0) {};
\node[name=b, fill=red, circle, double, minimum size=2, inner sep=0, outer sep=0, draw=red] at +(30:1) {};
\node[name=c, fill=red, circle, double, minimum size=2, inner sep=0, outer sep=0, draw=red] at +(90:1) {};
\node[name=d, fill=red, circle, double, minimum size=2, inner sep=0, outer sep=0, draw=red] at +(150:1) {};
\node[name=e, fill=red, circle, double, minimum size=2, inner sep=0, outer sep=0, draw=red] at +(210:1) {};
\node[name=f, fill=red, circle, double, minimum size=2, inner sep=0, outer sep=0, draw=red] at +(270:1) {};
\node[name=g, fill=red, circle, double, minimum size=2, inner sep=0, outer sep=0, draw=red] at +(330:1) {};
\draw[->,>=stealth',draw=red] (a) -- (b) node[right,red] at (b) {$J_{\text{\scalebox{0.7}{$1$}}}$};
\draw[<-,>=stealth',draw=red] (a) -- (c) node[above,red] at (c) {$J_{\text{\scalebox{0.7}{$2$}}}$};
\draw[->,>=stealth',draw=red] (a) -- (d) node[left,red] at (d) {$J_{\text{\scalebox{0.7}{$3$}}}$};
\draw[<-,>=stealth',draw=red] (a) -- (e) node[left,red] at (e) {$J_{\text{\scalebox{0.7}{$4$}}}$};
\draw[->,>=stealth',draw=red] (a) -- (f) node[below,red] at (f) {$J_{\text{\scalebox{0.7}{$5$}}}$};
\draw[<-,>=stealth',draw=red] (a) -- (g) node[right,red] at (g) {$J_{\text{\scalebox{0.7}{$6$}}}$};
\draw[->,>=stealth',draw=red] (b) -- (c);
\draw[<-,>=stealth',draw=red] (c) -- (d);
\draw[->,>=stealth',draw=red] (d) -- (e);
\draw[<-,>=stealth',draw=red] (e) -- (f);
\draw[->,>=stealth',draw=red] (f) -- (g);
\draw[<-,>=stealth',draw=red] (g) -- (b);
\node[right,red] at (a) {$J_{\text{\scalebox{0.7}{$0$}}}$};
\end{tikzpicture}
}

\usepackage{varioref}

\begin{document}

%\preprint{APS/123-QED}

\title{ String--Net Models with $\Z_N$ Fusion algebra}
% Force line breaks with \\

\author{Ling-Yan Hung}\email{jhung@perimeterinstitute.ca}
 \affiliation{Perimeter Institute for Theoretical Physics\\
              31 Caroline St N, Waterloo, ON N2L 2Y5, Canada}
\author{Yidun Wan}%
 \email{ywan@math.alice.kindai.ac.jp}
\affiliation{%
Research Center for Quantum computation, Kinki University\\
Kowakae 3-4-1, Higashi-Osaka 577-8502, Osaka, Japan
}%

\date{\today}% It is always \today, today,

\begin{abstract}
We study the Levin--Wen string--net model with a $\Z_N$ type fusion algebra. Solutions of the local constraints of this model correspond to $Z_N$ gauge theory and double Chern--Simons theories with quantum groups. For the first time, we explicitly construct a spin-$(N-1)/2$ model with $\Z_N$ gauge symmetry on a triangular lattice as an exact dual model of the string--net model with a $\Z_N$ type fusion
algebra on a honeycomb lattice. This exact duality exists only when the spins are coupled to a $\Z_N$ gauge field living on the links of the triangular lattice. The ungauged $\Z_N$ lattice spin models are a class of quantum systems that bear symmetry-protected topological phases that may be classified by the third cohomology group $H^3(\Z_N,U(1))$ of $\Z_N$. Our results apply also to any case where the fusion algebra is identified with a finite group algebra or a quantum group algebra.\end{abstract}

\pacs{11.15.-q, 71.10.-w, 05.30.Pr, 71.10.Hf, 02.10.Kn, 02.20.Uw}
\maketitle

\section{Introduction}\label{sec:Intro}
The classification of possible phases of matter is central to the study of condensed matter physics.
Until very recently different phases of matter have been associated with symmetry breaking,
which can be very succinctly described in the paradigm of Landau's effective theory.
Important leaps in our understanding come about when it is realized
that new phases of matter can arise even as
no symmetry breaking is involved.

It is therefore very important to give a systematic survey of these states of
matter, and ideally, provide a complete classification of them.

Very broadly speaking, gapped quantum phases of matter can be divided into two
classes: namely those involving long range entanglement (LRE) and those
involving only short range entanglement (SRE). When symmetries are present,
SRE displays a myriad of phases. For example Landau's paradigm of spontaneous symmetry
breaking belongs to the class of SRE. When symmetries are unbroken,
there are also distinct phases of matter, often termed
the symmetry protected topological (SPT) phases.  Their classification in terms
of group cohomology is recently given in Ref\cite{Chen2011d}.

On the other hand, the LRE phases of matter are
examples that realize \emph{topological order}, in which they display
features such as robust ground state degeneracies, non-Abelian statistics
of quasi--particle excitations, and
in many cases protected edge excitations. The classic examples of these phases include
the (fractional) quantum Hall states and chiral spin liquids. There is a very general
framework supplying exactly solvable models that incorporates
a large class of LRE phases, notably those preserving time-reversal symmetry.
This is called the string--net models \cite{Levin2004}, and it has been known that
the tensor category theory is the mathematical framework that underlies these models.

Very recently, a connection is discovered between a specific SPT  phase, namely an Ising spin model with  $\Z_2$ symmetry, and a LRE phase described by a string net model \cite{Burnell2012a, Gu2012}. %%JH 
In particular in the construction in \cite{Gu2012} it is found that when the $\Z_2$ symmetry of the spin model is gauged, it admits a dual description in terms of a string net model whose fusion rules are given exactly by $\Z_2$. It was conjectured that for a general SPT phase with discrete symmetry $G$, by gauging $G$ it admits a string net dual description with fusion rules also given by the product rule of $G$.

In this paper, by studying the explicit examples of string--net models with $\Z_N$ type fusion algebra, we construct
such a map between the string net models and the gauged SPT model. Although our
construction is based on $\Z_N$ fusion algebra, it is immediately applicable to
more general discrete groups $G$. This implies that the classification of SPT phases
provided by group cohomology in 2+1 dimensions via $H^3(G,U(1))$ described in Ref\cite{Chen2011d}
indirectly provide classifications of the corresponding string net models.
We support this claim also by studying the rescaling redundancy of the $6j$ symbols
that characterize a given string net model. We find that when the fusion rules
coincide with the product rule of a group $G$, the $6j$ symbols
can be interpreted as a 3-cocycle and that their rescaling redundancy can
be understood as an equivalence between these 3-cocycles up to a co-boundary
in the context of group cohomology\footnote{We were introduced to the mathematics literature
later and realize this is a well known fact. However, to our knowledge it has not
been pointed out explicitly in the context of string net models, and we believe there is
some value making the point more explicitly.}. Therefore these $6j$ symbols admit a classification
by $H^3(G,U(1))$, coinciding with that of the dual SPT phases.

Our paper is organized as follows. We begin in section II with a review of the basic ingredients
of the string net models. In section III we revisit the rescaling redundancy of the $6j$ symbols
and point out its relationship with group cohomology. In section IV, we study string--net models with $\Z_N$ type fusion algebra in greater details, and collect a number of useful facts about them. Some further details
and the explicit forms of $6j$ symbols corresponding to $\Z_N$ fusion algebra of various $N$ are
relegated to the appendix.
In section V, we construct the explicit map
between the string--net models with   $\Z_N$ fusion algebra and the corresponding gauged SPT model, generalizing
the construction proposed in \cite{Levin2012}. We note that the relationship between these SPT phases
and the string--net models are explored via a different route also in \cite{Burnell2011a}. %%JH

Finally we conclude in section VI and point to several open problems.

%We got to say that this duality enables us to classify the SPT  phases, which are the zn spin models, by means of the cohomology group of $\Z_N$.

\section{Review of the fundamentals}\label{sec:Rev}
In this section, we review the fundamentals of string--net models by collecting the basic ingredients, in the hope that
those who are new to this area may still find this paper accessible.

%\subsection{String--net models}\label{subsec:revSNmodel}

Levin--Wen string--net model is known as the Hamiltonian formulation of the Turaev--Viro model\cite{Hu2012,Wang2010,Turaev1994,KADAR2009}. As such, at the level of ground states, these two models are equivalent. String--net models are usually defined on the honeycomb lattice, which is also the convention we take in this paper. A string--net model is actually a family of models characterized by a triple $\{H,N,\F_N\}$, where $H$ is the Hamiltonian, $N$ the number of string types, which does not count the trivial string type $0$ that is always  present in the model, and $\F_N$ is the fusion algebra over the $(N+1)$ string types. 
The Hamiltonian is supposed to be exactly soluble. First of all it enforces the fusion rules $\F_N$ dynamically so that states containing
vertices where different string types meet and yet violate $\F_N$ are separated from those satisfying $\F_N$ by a large gap. 
Such $\F_N$ violating vertices are thus entirely excluded in the low energy limit. 
Secondly the Hamiltonian favors its ground state, albeit with a smaller gap, wave functions that can be interpreted as the fixed--point wave functions of certain IR renormalization flow. %JH
Hence, these wave functions appear to be the same at all length scales and as such must satisfy certain local constraints, which are to be presented shortly. We shall delay further discussion of the Hamiltonian but dwell on the other two elements in the tuple first.

String degrees of freedom reside on the edges of the honeycomb lattice. In the most general setting, the $N+1$ string types are merely abstract numbers. In the cases that have been so far studies, they either label the group elements of certain finite group\cite{Kitaev2006} %%JH
or the irreducible representations of a finite (quantum) group\cite{Levin2004}. %%JH
In the former case, the string type $0$ labels the identity of the group, while in the latter the trivial representation. Generally speaking, the input data of a string--net model is the fusion  algebra $\F_N$ of the $N+1$ string types, denoted by $0,1,\dots,i,\dots, N$, which takes the following form.
\be\label{eq:fusionAlg}
i\otimes j=\sum\nolimits_k \N^k_{ij} k\, ,
\ee
where $\N^k_{ij}$ is the multiplicity of the string type $k$ that appears in the direct--sum decomposition of the tensor product $i\otimes j$. 
If we define  $\N_{ijk}=\N^{k^*}_{ij}$, %JH
where $k^*$ denotes the conjugate of string-type $k$,
then $\N_{ijk}$ is the number of occurrences of the trivial string type in the product $i\otimes j\otimes k$. Note that the fusion algebra is in general non--commutative if it is  a group algebra but commutative if its elements are representations of a (quantum) group.

In this paper, however, we focus on the cases where the string types label the irreducible representations of certain finite group or quantum group. A situation often encountered is that the fusion algebra  coefficients satisfy $\N_{ijk}=\delta_{ijk}$, where $\delta_{ijk}=1$ if $i\otimes j\otimes k=0$ and otherwise $\delta_{ijk}=0$, such that the fusion algebra can be identified with certain Abelian groups. 
We should emphasize that this is true for Abelian groups, but not in general. The fusion algebra of $\Z_N$ type falls in this class, and is what we shall consider in this paper from now on.  %JH
As such, there exists a unique notion of conjugate string type: If $i\otimes j=0$, we say that $j$ is the conjugate string type of $i$ and write $j=i^*$. The string type $i$ is called self--conjugate if $i=i^*$.

\begin{figure}[h!]
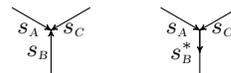

\centering
   \Ygraph{s\s{a}}{s\s{b}}{s\s{c}}\qquad \Ygraph[0]{s\s{a}}{s^*\s{b}}{s\s{c}}
\caption{A string--net vertex.}\label{fig:SNvertices}
\label{fig:snVertex}
\end{figure}

We now visualize the fusion algebra $i\otimes j=k^*$ on the honeycomb lattice. Fig. \ref{fig:snVertex} shows a vertex of the honeycomb lattice. An edge $A$ of the lattice is graced with a string type $s\s{a}$ and endowed with an orientation, specified by an arrow. Such a string of type $s\s{a}$  can also be represented by a flipped arrow, but with the conjugate string type
$s\s{a}^*$.

The string--net wave function satisfies the local constraints listed below.
\bse\label{eq:SNwaveFunc}
\begin{align}
\Phi\bpm\shadowGraph\hspace{-0.3pt}\sline\hspace{-1pt}\shadowGraph\epm
&=\Phi\bpm\shadowGraph\hspace{-0.3pt}\curveline\hspace{-1pt}\shadowGraph\epm
\label{eq:SNwaveFuncDeform}\\
\Phi\bpm\shadowGraph & \loopGraph{i}\epm
&=d_i\Phi\bpm\shadowGraph\epm\label{eq:SNwaveFuncLoop}\\
\Phi\left(\bmm\vspace{-3pt}\bubbleGraph{i}{j}{k}{l}\emm\right)
&=\delta_{ij}\Phi\left(\bmm\vspace{-3pt}\bubbleGraph{i}{i}{k}{l}\emm\right)
\label{eq:SNwaveFuncBubble}\\
\Phi\left(\bmm\vspace{-3pt}\scalefont{0.6}\Hgraph{j}{m}\emm\right)
&=\sum\limits_nF^{j_1j_2m}_{j_3j_4n}\Phi\left(\bmm\vspace{-3pt}\scalefont{0.6}\Xgraph{j}{n}\emm\right),
\label{eq:SNWaveFuncCross}
\end{align}
\ese
where the gray blocks in the first two rows represent the rest of the string--net, and all other graphs in the wave functions are understood as subgraphs of the string net. More accurately speaking, $\Phi$ 
is the weight of the configuration in the ground-state
wave function of the Hamiltonian H. However we will keep to the terminology in \cite{Levin2004} and refer to $\Phi$ loosely
as the wavefunction. %JH
The motivations and meanings of these local rules are in order: Eq. (\ref{eq:SNwaveFuncDeform}) reads that the string--net wave function is invariant under local, continuous deformation; Eq. (\ref{eq:SNwaveFuncLoop}) implies that a loop disconnected from the string net contributes only a scaling constant---the quantum dimension---to the wave function. Eq. (\ref{eq:SNwaveFuncBubble}) is a motivated by the physical intuition that the bubble is not observable at large scale; Eq. (\ref{eq:SNWaveFuncCross}) is the crossing symmetry , which is an ansatz motivated by Conformal Field Theory. Another property of a string--net wave function is that it is invariant under the addition of an edge with string type $s=0$ that connects any two edges in the string--net. This property and Eq. (\ref{eq:SNwaveFuncBubble}) implies that any tadpole in a string--net is equivalent to the case on the LHS of Eq. (\ref{eq:SNwaveFuncLoop}).

 The $6j$ symbols $F^{ijm}_{kln}$ satisfy the following relations.
\be\label{eq:F6jRel}
\begin{aligned}
& F^{ijm}_{j^*i^*0}=\frac{v_m}{v_iv_j}\delta_{ijm}\\
& F^{ijm}_{kln}=F^{klm^*}_{ijn^*}=F^{jim}_{lkn^*}=F^{mij}_{nk^*l^*}\frac{v_mv_n}{v_jv_l}\\
& \sum\limits_n F^{mlq}_{kp^*n}F^{jip}_{mns^*}F^{js^*n}_{lkr^*}=F^{jip}_{q^*kr^*}F^{riq^*}_{mls^*}.
\end{aligned}
\ee
and $v_i^2 = d_i$. %JH
We refer interested reader to \cite{Turaev1994} for a more detailed introduction
to the $6j$ symbols and their place in representation theory. %JH
The second and third relations are the tetrahedral symmetry, and the pentagon identity respectively. The first relation comes from a rescaling redundancy of the wave function which we shall explain in more detail in the next sub--section.

Equations (\ref{eq:F6jRel}) may admit many solutions. Each such solution up to the rescaling redundancy to be elaborated in Section \ref{sec:phasered} is believed to yield a distinct string--net model in the family specified by $\{H,N,\F_N\}$. We thus denote a single string--net model by the quadruple $\{H,N,\F_N,[F]\}$, where $[F]$ is set of equivalent solutions to Eqs. (\ref{eq:F6jRel}).

We also note that in the third equality in the tetrahedral symmetry relations, the symmetry involves swapping the first two columns of the 6j-symbols descends from the reflection symmetry of the tetrahedron. It has been argued that this relation might be too strong and potentially exclude physically viable and interesting solutions. Nonetheless, we will in this paper for simplicity retain this as part of the symmetry of the $6j$ symbols.

It is often convenient to do computations in terms of the symmetric $6j$ symbols $G^{ijm}_{kln}$, defined by
\be\label{eq:F2G6j}
F^{ijm}_{kln}=G^{ijm}_{kln}v_m v_n,
\ee
because $G^{ijm}_{kln}$ is manifestly symmetric under tetrahedral transformation, seen as follows along with two other relations corresponding to those in Eq. (\ref{eq:F6jRel}).
\bse\label{eq:G6jRel}
\begin{align}
& G^{ijm}_{j^*i^*0}=\frac{\delta_{ijm}}{v_iv_j},\ \ G^{ii^*0}_{m^*mj}=\frac{\delta_{ijm}}{v_iv_m}\label{eq:G6jRelNor}\\
& G^{ijm}_{kln}=G^{klm^*}_{ijn^*}=G^{jim}_{lkn^*}=G^{mij}_{nk^*l^*}\label{eq:G6jRelTet}\\
& \sum\limits_n d_nG^{mlq}_{kp^*n}G^{jip}_{mns^*}G^{js^*n}_{lkr^*}=G^{jip}_{q^*kr^*}G^{riq^*}_{mls^*}\label{eq:G6jRelPen}.
\end{align}
\ese
Taking values $r=0$, $r=l$, and $j=k^*$, Eq. \ref{eq:G6jRelPen} leads to an orthogonality relation (also called a $2G$ relation):
\be\label{eq:2Grel}
\sum\limits_n G^{mlq}_{kp^*n}G^{l^*m^*i^*}_{pk^*n}d_n=\frac{\delta_{iq}}{d_i}\delta_{mlq}\delta_{k^*ip}.
\ee
Hereafter, we shall simply refer to the $6j$ symbols $F^{ijm}_{kln}$ as the $F$--symbols and the symmetric $6j$ symbols $G^{ijm}_{kln}$ the $G$--symbols.

The $6j$ symbols are the necessary ingredients that define
the action of the magnetic flux operators $B_p^s$, which in turn are building
blocks of the Hamiltonian.

Since we will be primarily focusing on the models with $\Z_N$ fusion algebra, we postpone the introduction of the magnetic flux operators and the string--net Hamiltonian for later  after a close look at the $6j$ symbols of $\Z_N$ type fusion algebra in Section \ref{sec:ZnSN}. For our convenience, we shall denote the $\Z_N$ type fusion algebra by $\F_{\Z_N}$.
\section{A note on group cohomology classification of $6j$-symbols}\label{sec:phasered}
It was noted in Ref\cite{Levin2003} that there is a rescaling symmetry that preserves
the pentagon identity. The idea is that for each given wave function,
a rescaling of each of the vertices $A$ by a phase factor $f_{i_A\,j_A\,k_A}$,
where $i_A,j_A,k_A$ are the three in-going string states connected to the vertex,
does not lead to new physics. In other words,
the rescaling that takes $\Phi$ to $\tilde{\Phi}$, i.e.,
\be
\Phi \to \tilde\Phi = \prod_{A} f_{i_A,j_A,k_A} \Phi,
\ee
where $A$ denotes the vertices of the string--net that $\Phi$ describes,
is a redundancy of the model.

The corresponding $\tilde{F}$ symbols that describe crossing relations between
the $\tilde{\Phi}$ (see Eq. (\ref{eq:SNWaveFuncCross})) is thus related to the original $F$ symbols by
\be\label{eq:phase1}
\tilde{F}^{ijm}_{kln} = F^{ijl}_{klm} \frac{f_{n^*jk}f_{nli}}{f_{ijm}f_{klm^*}}.
\ee
One can easily check that this rescaling preserves the pentagon identity, provided
that $f_{ijk}$ is symmetric under cyclic rotation of the indices.

Let us pause here and remark about the connection of the rescaling redundancy
with group cohomology. When the fusion rules are such that the product
of irreps forms a group $G$, the fusion rules dictate that at a non-vanishing
vertex $A$ where the string states satisfy $i_A \otimes j_A \otimes k_A = 0$. %JH
Therefore,
one can think of $k_A$ as determined by $i_A$ and $j_A$, and thus $f_{i_Aj_Ak_A}$
is a map $f:G^2 \to U(1)$.
Similarly, there are three independent constraints between the 6 indices
of $F$. As a result, there are exactly three independent indices.
Hence, $F$  is in fact a map $F:G^3 \to U(1)$. In this light, the pentagon identity can be viewed as the statement that
$F$ is a 3-cocyle. Such an interpretation is already well known
in the study of tensor category theory. The rescaling redundancy
involving the product of four $f_{i_A,j_A,k_A}$ comes precisely in the form
of an 3-coboundary.  As a result, inequivalent $F$-symbols are classified
by $H^3(G,U(1))$.
We note however that our examples of $6j$ symbols detailed in the appendix
do not span $H^3(G,U(1))$.
Specifically, it is known that $H^3(\Z_N,U(1)) = \Z_N$ in two dimensions \cite{Chen2011d},
but in the case
of $\Z_3$ for example, we found only one distinct solutions to the pentagon relations.
We believe this is a result of our assumption of the tetrahedral symmetry
which has excluded some of the viable solutions. In the case where $d_i=+1$ this
is identical to the standard $6j$ symbols of the $Z_3$ group, as detailed for example
in the appendix in Ref. \cite{Bonderson2007}. %%JH

Due to our assumption of the reflection symmetry of the tetrahedron, the choice of normalization that
fixes the rescaling redundancy above %JH
implies further that $f_{ijk}$ is symmetric under exchange of \emph{any}
two indices.
To distinguish physically different states, it is therefore convenient
to fix this redundancy.
In Ref\cite{Levin2004} a choice is made such that $f_{ijk}$ is tied to the
wave function of the $\theta$ graph, i.e.,
\be\label{eq:gauge1}
f_{ijk}f_{i^*k^*j^*} = \frac{v_{i} v_{j} v_{k}\Phi(\emptyset)}{\Phi\left(\bmm\scalebox{0.6}\thetaGraph\emm\right)}.
\ee

Using this relation, and that
\be
\Phi\left(\bmm\scalebox{0.8}\thetaGraph\emm\right) = F^{ijk}_{j^*i^*0} d_i d_j \Phi(\emptyset),
\ee
we arrive at the normalization condition given in the first equation in (\ref{eq:F6jRel}).
We note, however, that the gauge condition fixes only the value of the product $f_{ijk}f_{i^*k^*j^*}$.
This leaves us with further freedom to rescale $f_{ijk}$, as long as the rescaling is
absorbed by $f_{i^*k^*j^*}$:
\be
f_{ijk}\to f_{ijk} e^{\ii \vartheta}\,, \qquad f_{i^*k^*j^*}\to f_{i^*k^*j^*} e^{-\ii \vartheta}.
\ee

We will demonstrate in the explicit example of $\Z_6$
how the residual redundancy should be fixed.
\section{$\F_{\Z_N}$ String--Net Models}\label{sec:ZnSN}

In the following, we will focus on the $\F_{\Z_N}$ string--net models defined on the honeycomb lattice.
As reviewed in the previous section, the string degrees of freedom reside on the edges of some lattice.
In a $\F_{\Z_{N}}$ model there are $N$ string types, including type $0$, that could live on each edge. These string types
are elements of the fusion algebra $\F_{\Z_N}$. Since $\F_{\Z_{N}}$ is isomorphic to the group $\Z_{N}$, we can replace the tensor product symbol $\otimes$ in the algebra simply by the operator $+$, which is the product of $\Z_{N}$ group elements. As such, the conjugate of a string type $i$ can be written as $i^*$ or equally as $-i$. Recall that each edge $A$ is endowed
with an orientation, specified by an arrow. In the case of $\F_{\Z_N}$, therefore,
each edge of the lattice could take any integer values modulo $N$ with a chosen orientation.
If the orientation is flipped,
we have to flip the sign of the value the edge takes, i.e., sending $s\s{A} \to -s\s{A}=s\s{A}^*$.
As in the previous section, we need to specify the branching rules
satisfied by string states residing on edges that meet at a vertex.
When all the orientations of the meeting edges are chosen to be pointing
toward the vertex, then the branching rule is given by
\be
\Ygraph{s}{i}{j}\, \qquad s+i+j=0 \pmod{N}.
\ee

The quantum dimensions and $6j$-symbols relevant for
defining local rules and crossing symmetry of the model is studied
in the following sub--section.

\subsection{$\F_{\Z_N}$ $6j$-symbols and quantum dimensions: General Properties}\label{subsec:zn6jQDim}
To facilitate constructing and gain better understanding of $\F_{\Z_N}$ string--net models, we would like to study the general properties of the $\F_{\Z_N}$ $6j$ symbols and quantum dimensions, which are the fundamental building blocks of the models.

To begin with, we
recall that the $6j$ symbols are related by the tetrahedral symmetry as given in Eq. (\ref{eq:F6jRel}) and in Eq. (\ref{eq:G6jRelTet}).
There is a simple rule to determine if two $6j$ symbols are not related by tetrahedral symmetry.
By noting that the number of occurrences of a self-conjugate string type $s=s^*$ in the $G$--symbol (and thus the $F$--symbol)
is an invariant of the tetrahedral symmetry,
we conclude that any two $6j$-symbols which contain different numbers of
self-conjugate string types must not be related by the tetrahedral symmetry.

There are several important properties one could derive using the $2G$ relation Eq. (\ref{eq:2Grel}).
%We consider the case where the RHS of Eq. (\ref{eq:2Grel}) is nonzero.
Non-vanishing $6j$ symbols are those that satisfy branching rules.
To obtain nontrivial constraints following from Eq. (\ref{eq:2Grel}),
$i=q$, $m+l+q=0\pmod{N}$, and $k^*+i+p=0\pmod{N}$ always hold;
Note also that in the case of $\Z_N$, for any given $l$ and $p$, $\delta_{l^*n(p+i)^*}= 1$ has a unique
solution --- $n=l+p+i\pmod{N}$ ---  that gives rise to the only non--vanishing term in Eq. (\ref{eq:2Grel}), which,
by substituting  $m=i^*+l^*\pmod{N}$, $k=p+i\pmod{N}$ and $n=l+p+i\pmod{N}$ in the LHS, turns out to be
\be\label{eq:2GrelRHSis1}
G^{(l+i)^*l\,\,\,i}_{(p+i)\,p^*\,(l+p+i)}G^{l^*\,(l+i)\,\,\,i^*}_{p\,(p+i)^*\,(l+p+i)}d_{l+p+i}\,d_i=1.
\ee
To avoid clutter,
$\pmod{N}$ has been omitted but implicitly assumed in all
arithmetic of string types appearing in the $6j$ symbols above and in any subsequent
symbols.

Consider the special case $i=0$ in Eq. (\ref{eq:2GrelRHSis1}):
\[
\left(G^{l^*l\,\,\,0}_{pp^*\,(l+p)}\right)^2 d_{l+p}\xeq[]{Eq. (\ref{eq:G6jRelNor})}\frac{d_{l+p}}{v^2_l v^2_p}=1.
\]
We thus arrive at
a \textbf{product rule of quantum dimensions}. Namely,
\be\label{eq:3qDim}
d_{l+p}=d_l d_p
\ee
where $d_x=v_x^2$ is applied. There exists a generalization of this product rule to the case where $\N_{ijk}\neq\delta_{ijk}$, which reads
\[
d_i d_j=\sum_k \N^k_{ij}d_k.\]
We do not dwell on this generalization but invite one to the reference\cite{Kitaev2006},
and for a discussion of $\Z_N$ and specifically $\Z_3$ in \cite{Tagliacozzo2012,Schulz2012}. %JH
An immediate consequence
of Eqn. (\ref{eq:3qDim}) is that taking $l=p^*$ gives:
\be\label{eq:qDimSquare1}
d_l^2\equiv 1\quad\forall l\in\{0,1,\dots,N-1\}.
\ee
Hence, all $\F_{\Z_N}$ quantum dimensions are either $+1$ or $-1$.

Eq. (\ref{eq:3qDim}) has more information to be extracted.
Suppose we demand that $l+p=l^*\pmod{N}$ in Eq. (\ref{eq:3qDim}), then it follows that
\be\label{eq:qDimEven1}
d_{p}\equiv 1\,,\qquad 2l^*=p\pmod{N}
\ee
In other words $d_p=1$ if $p$ is \emph{even} in modular arithmetic.
For even $N$, \emph{even} numbers take on the same meaning as usual.
On the other hand, since
$q = q-N \pmod{N}$ for any $q$,
$q-N$ is even for any odd $q$ when $N$ is odd.
As a result, \emph{all} elements
$p$ of $\Z_N$ satisfy the condition $p= 2l^* \pmod{N}$ for some
$l \in  \{0, \cdots N-1\}$ when $N$ is odd.

When $N$ is even, odd numbers in $\Z_N$ are ambiguously defined; hence, in this case, quantum dimensions $d_i$ with odd $i$ allow two possible values, $\pm 1$. But fortunately, there is no freedom of making independent choices of the signs of these quantum dimensions. The reason is quite obvious, as any two odd numbers, say, two neighbouring ones $2k+1$ and $2k-1$, differ by an even number, then by the product rule Eq. (\ref{eq:3qDim}), we have
\[
d_{2k+1}=d_{2k-1}d_2\equiv d_{2k-1},
\]
from which we infer that in $\Z_{N\in 2\Z}$, $d_i=d_j$, for all $i,j\in 2\Z_N+1$. Therefore, we end up with at most two overall choices of the signs of the quantum dimensions that take value in $\{\pm 1\}$.

We summarize these important and general results in the following theorem.
\begin{theorem}
For a $\F_{\Z_N}$ string--net model, if $N\in 2\Z+1$, all the quantum dimensions $d_i$
of the model satisfy $d_i\equiv1$.
If $N\in 2\Z$, $d_i \equiv 1$ for $i\in 2\Z_N$, and $d_j=d_k$, $\forall j,k\in 2\Z_N+1$.
\end{theorem}
Thus, only in $\F_{\Z_N}$ string--net models with $N\in 2\Z$, there can exist quantum dimensions with a minus sign.
Clearly, for any $d_i\equiv1$, the corresponding $v_i=\pm1$, whereas for any $d_i=\pm1$, the corresponding $v_i$ can take values in $\{\pm1,\pm\ii\}$.

Let us  take a closer look at Eq. (\ref{eq:2GrelRHSis1}). By means of the tetrahedral symmetry of $G$--symbols, namely Eq. (\ref{eq:G6jRelTet}), we can turn Eq. (\ref{eq:2GrelRHSis1}) into a nicer form as
\be\label{eq:2Gunity}
G^{n^*p(l+i)}_{il(p+i)}G^{np^*(l+i)^*}_{i^*l^*(p+i)^*}d_nd_i=1,
\ee
where we recall that $n= l+p+i \pmod{N}$.
The uniqueness of the number $n$ in Eq. (\ref{eq:2Gunity}) guarantees that the $G$--symbol appearing in Eq. (\ref{eq:2Gunity}) can be considered as the \textbf{canonical forms} of $\F_{\Z_N}$ $G$--symbols, in the sense that any non-vanishing $\F_{\Z_N}$ $G$--symbol
can be casted (by the tetrahedral symmetry)
in the form of them. (Note that the $G$ sitting on the right above has the same form as the first except that all string types are conjugated.) To see the significance of Eq. (\ref{eq:2Gunity}) and the canonical forms of $G$--symbols, we rewrite Eq. (\ref{eq:2Gunity}) in terms of the $F$--symbols by Eq. (\ref{eq:F2G6j}) and obtain
\[
F^{n^*p(l+i)}_{il(p+i)}F^{np^*(l+i)^*}_{i^*l^*(p+i)^*}d^{-1}_{i+l}d^{-1}_{p+i}d_nd_i = 1. %JH
\]
But by $n= l+p+i \pmod{N}$ and Eqs. (\ref{eq:3qDim})  and (\ref{eq:qDimSquare1}),
\[
\begin{aligned}
d^{-1}_{i+l}d^{-1}_{p+i}d_nd_i = d_l^2 d_p^2 \equiv 1.
\end{aligned}
\]
Therefore, the following identity holds for $\F_{\Z_N}$ $F$--symbols.
\be\label{eq:2Funity}
F^{n^*p(l+i)}_{i\,\,\,\,l\,\,\,(p+i)}F^{n\,p^*(l+i)^*}_{i^*l^*(p+i)^*}\equiv 1.
\ee
Note that $i,p,l$ are completely independent, and that $n$ was fixed
only from branching rules. Therefore, either of the $F$--symbols
in Eq. (\ref{eq:2Funity}) above
is generic.
Furthermore, according to Ref\cite{Levin2004}, if we require the
Hamiltonian of a string--net model to be Hermitian, the $F$--symbols of the
model must satisfy $\bigl(F^{ijm}_{kln}\bigr)^*=F^{i^*j^*m^*}_{k^*l^*n^*}$,
in which case Eq. (\ref{eq:2Funity}) becomes
\be\label{eq:FsymbNorm1}
\left|F^{n^*p\,(l+i)}_{i\,\,\,\,l\,\,(p+i)}\right|^2\equiv 1.
\ee

From now on, we shall  consider throughout this paper that the Hamiltonian of a $\F_{\Z_N}$ string--net model is Hermitian.

We designate the $F$--symbol in Eq. (\ref{eq:2Funity}) the \textbf{canonical form} of $\F_{\Z_N}$ $F$--symbols. One should note from the last equality in the second row of Eq. (\ref{eq:F6jRel}) that a non-canonical $F$--symbol may differ from its canonical form by a phase factor. Fortunately, in the case of $\F_{\Z_N}$, the phase factor is but a sign, as we now show. Take the canonical $F$--symbol in Eq. (\ref{eq:FsymbNorm1})
and act on it with the tetrahedral transformation that brings the phase factor:
\[
F^{n^*p\,(l+i)}_{i\,\,\,\,l\,\,(p+i)}=F^{(l+i)n^*p}_{(p+i)i^*l^*}\dfrac{v_{l+i}v_{p+i}}{v_p v_l}.
\]
Since $v_x=\pm \sqrt{d_x}$, the phase factor is actually
\be\label{eq:znFtetsign}
\dfrac{v_{l+i}v_{p+i}}{v_p v_l}=\pm\sqrt{\dfrac{d_{l+i}d_{p+i}}{d_p d_l}}\xeq[]{Eq. (\ref{eq:3qDim})}\pm\sqrt{d_{2i}}=\pm 1,
\ee
which is real and merely a sign. Therefore, all of $\F_{\Z_N}$ $F$--symbols strictly satisfy Eq. (\ref{eq:FsymbNorm1}), and thus live on the unit circle on the complex plane.
In this section we are not going to compute the $6j$ symbols explicitly but leave it for Appendix \ref{appx:6jSymbols}, where one will see that
some $F$--symbols are identical to one while some others admit more than one solutions, namely either $\{\pm 1\}$; the rest admit still more
possibilities which are in general complex and constrained only to have norm one.

\subsection{The Magnetic Flux Operators}\label{subsec:ZnStringOp}
Given the knowledge of the $\F_{\Z_N}$ $6j$ symbols acquired in the previous section and in Appendix \ref{appx:6jSymbols}, we are ready to construct and understand the magnetic flux operators $B_p$ defined on a hexagonal plaquette $p$ of the honeycomb lattice.
\be
\begin{aligned}
B^s_p &\BLvert\Psix[1]{l}{e}\Brangle
= \sum\limits_{e'\s{a}}
F^{l_1 e^*_1 e_6}_{s^* e'_6 e'^*_1}F^{l_2 e^*_2 e_1}_{s^* e'_1 e'^*_2}F^{l_3 e^*_3 e_2}_{s^* e'_2 e'^*_3} \times\\
&F^{l_4 e^*_4 e_3}_{s^* e'_3 e'^*_4}F^{l_5 e^*_5 e_4}_{s^* e'_4 e'^*_5}F^{l_6 e^*_6 e_5}_{s^* e'_5 e'^*_6} \BLvert\Psix[1]{l}{e'}\Brangle
\end{aligned},
\ee
where the summation runs over all primed string types $e'_1$ through $e'_6$. In fact, in the case of $\F_{\Z_N}$, only one term survives the summation, as demanded by the branching rule, $s*+e'\s{a}+e^*\s{a}=0\pmod{N}$, which is pointed out in the previous subsection. In particular, according to the canonical form in Eq. (\ref{eq:2FunityNew}) of the $F$--symbols in the summation above, the matrix elements are already in their canonical forms; hence, the nonzero matrix elements of the operator $B^s_p$ can be written as
\be
\begin{aligned}
B^{s,e'_1e'_2e'_3e'_4e'_5e'_6}_{p,e_1e_2e_3e_4e_5e_6}(l_1,l_2,\dots,&l_6)_{e'\s{a}=e\s{a}+s}\\ &=\prod\limits_{A=1}^{6}F^{l\s{a}e^*\s{a}e\s[-1]{a}}_{s^*(e\s[-1]{a}+s)(e\s{a}+s)^*}\;,
\end{aligned}
\ee
where the addition is defined in modulo--$N$ arithmetic. One immediately sees that the action of a $B^s_p$ on a plaquette does not alter the legs of the plaquette but simply shifts the string type of every edge of the plaquette up by $s\pmod{N}$.
In general, according to Eqs. (\ref{eq:2FunityNew}) and (\ref{eq:znFasPhase}), all $\F_{\Z_N}$ $F$--symbols are complex numbers with norm one and can be written as $F^{l\s{a}e^*\s{a}e\s[-1]{a}}_{s^*(e\s[-1]{a}+s)(e\s{a}+s)^*} =\e^{i\alpha\s{n}(s,l\s{a},e\s{a})}$.

Let us also introduce operators $\Sigma^{\pm n}\s{A}$ that are responsible
for cyclically shifting the string type of an edge $A$  by $ n$--units, $n\in\Z_N$,  by the modular arithmetic $e\s{a}+n\pmod{N}$.
The only non--vanishing components are given by
\be\label{eq:stringShiftOp}
(\Sigma^{n}\s{a})_{\kappa \lambda} = 1, \,\, \lambda = \kappa +n \pmod{N}\,.
\ee
In particular, $\Sigma^{0}\s{a}\equiv 1$. Our definition ensures that $\Sigma^{n+N}\s{a} = \Sigma^{n}\s{a}$.
Note that shift operators commute with each other, and that
\be
\Sigma^{n}\s{a}\Sigma^{m}\s{a} = \Sigma^{n+m}\s{a}.
\ee
In other words they also satisfy modular arithmetic.
These operators can be constructed simply using the raising
and lowering operators of $SU(2)$.

Therefore, we can write the operator form of a generic $B^s_p$ as
\be\label{eq:znBpsGeneral}
B^s_p(l_1,l_2,\dots,l_6)=\exp(\ii\mathsmaller\sum_{\mathsmaller {A\in p}}\alpha\s{n}(s,l\s{a},e\s{a}))\prod\limits_{A\in p}\Sigma^{s}\s{a}\,,
\ee
where $A\in p$ means all six edges of the plaquette $p$.
In the case where $s=0$, it is obvious that
\be
B^0_p\equiv 1\,.
\ee

\subsection{The Hamiltonian}
The usual string--net Hamiltonian takes the form
\be\label{eq:snH}
H=-\sum_v A_v-\sum_p B_p,
\ee
where $A_v\Bigl\vert\bmm\scalebox{0.6}{\Ygraph{i}{j}{k}}\emm\Bigr\rangle =\delta_{ijk}\Bigl\vert\bmm\scalebox{0.6}{\Ygraph{i}{j}{k}}\emm\Bigr\rangle$ is the vertex operator defined at each string--net vertex, and $B_p=\sum_{s=0}^N a_s B^s_p$, with $a_s=d_s/D$ and $D=\sum_{i=0}^N d^2_i$, is the "magnetic--flux" operator defined for each hexagonal plaquette of the string--net lattice. Each operator $B^s_p$ in $B_p$ is defined in the previous sub--section. The vertex operators $A_v$ are projectors. The parameter $a_s$ ensures that the operators $B_p$ are also projectors. It can be shown that $\{A_v,B_p|\forall v,p\}$ is a set of commuting operators, whose common eigenstates span the Hilbert space of the model. The ground states of the model are thus the $+1$ eigenstates of $A_v$ and $B_p$, which are known as the string--net condensed states, characterized by the local rules Eq. (\ref{eq:SNwaveFunc}).

\section{The duality between $Z_{N}$ string--net models and (SPT) spin models}
In the following, we will explain in detail the duality between general $\F_{\Z_N}$
string--net models on the honeycomb lattice and spin--$(N-1)/2$ models on
the triangular lattice. In the special case where $N=2$,
the duality has been explained in detail in Ref\cite{Levin2012,Burnell2011}. %JH
 As we will explain, however,
for $N>2$, there are several new elements that are needed.
Therefore, we will first begin with an explanation of the general case before illustrating these
general principles in greater detail in specific examples.

There are several elements in a string net model that are mapped to the spin model. We
will deal with them in several steps in the following sub--sections.

\subsubsection{Turning the strings states into spin states}

The duality between a string--net model and a (gauged) spin model is one  that
places the string--net model  on a honeycomb lattice and the spin model on the dual
triangular lattice.
As usual, the string states live on the edges, while the spin states sit on the vertices.
By comparing these two models,
however, we realize that there is some redundancy in the string net model in specifying
both the string type on and the orientation of an edge. Such a redundancy---in particular the orientation---is necessary in describing the crossing symmetry. While
we note that the crossing symmetry is indeed a symmetry that preserves
the dimension of the Hilbert space \cite{Hu2012}, as the intermediate string type $n$ in Eq. (\ref{eq:SNWaveFuncCross})
is summed over, it clearly alters the lattice structure. The duality we
study here is otherwise defined only for a given fixed lattice structure on
both sides of the map. In particular, crossing symmetry necessarily violates
the valency of the dual triangular lattice, rendering the duality map ill--defined.
A fixed lattice structure on the other hand, is also
the situation that is most physically relevant, since any lattices
pertinent to experiments are in reality solids whose structure is essentially fixed below
their melting points.

Having come to terms with a fixed lattice, thereby giving up the freedom
to deform the honeycomb lattice on which the string--net model
concerned is defined via crossing, we are then justified in
choosing a convention for orienting each edge. Such removal
of the orientation redundancy would render the duality map
with the spin model most transparent.

One natural way to fixing orientation is to make use of the
fact that the honeycomb lattice can be divided into two sub--lattices, $L_{BL}$
and $L_{WH}$, whose sites are colored black and white respectively in Fig. \ref{fig:Duality}.
Since edges always connect sites of opposite colors, we can uniquely
fix the orientation of
each edge by requiring that the arrow always
points toward a black vertex. Having fixed the orientation
globally,
we can then assign a unique string type to each edge, which
can in turn be interpreted as a spin state.
As such, we can remove the arrows on the edges of the honeycomb lattice. In the following we will explain how these spin states can
be translated into spin states defined on a triangular lattice.
\begin{figure}[h!]
\centering
\dualityGraph
\caption{Duality}\label{fig:Duality}
\end{figure}
\subsubsection{Branching rules and the dual spin model on a triangular lattice}
The Hamiltonian
is chosen to favor energetically the states satisfying the branching rules as specified
in the previous section corresponding to the fusion algebra $\F_{\Z_N}$
at the vertices. These branching rules restrict the string types on the three edges that meet
at a vertex so that they have to add up to a trivial representation.
Now having already fixed the orientation using the convention specified in the previous subsection,
the branching rule for $\F_{\Z_N}$ can be stated unambiguously as follows:
that the values of the string types
on the three edges that meet at a vertex must sum up to zero modulo $N$, without
further reference to orientations.
We emphasize again that on the orientation--fixed string--net, crossing
symmetry transformations cannot be applied.

Now it is time to introduce the relation between the string--net model and the dual spin model
defined on a triangular lattice. Particularly, we would like to understand
how the duality map relates the Hilbert spaces of the two models.
The triangular lattice is the dual lattice of the honeycomb lattice such that each of its
vertex resides at the center of a hexagon in the honeycomb lattice (see Fig. \ref{fig:Duality}). A spin degree of freedom lives on each vertex of the triangular lattice.
Two spins $J_a$ and $J_b$ are adjacent if they are right on the two neighbouring vertices $a$ and $b$ connected by a straight line, which necessarily crosses (perpendicularly) an edge
(of string type $s\s{a}$) of the honeycomb lattice.
Now we assign an orientation to this straight line by picking an arrow pointing toward $b$, and
dictate that the duality map is given by
\be\label{eq:dualmap1}
s\s{a} = J_a - J_b \pmod{N}.
\ee
The choice of orientation on different edges of the triangular lattice is not independent.
In fact, it has to be chosen such that on every triangle the orientations of the three edges form a closed loop. Since neighboring triangles share an edge, they have opposite orientations. At the center of each triangle is a vertex of the honeycomb lattice. Now it should be clear that the map Eq. (\ref{eq:dualmap1}) ensures that for an arbitrary set of spins on the triangular lattice, the resulting dual set of spins defined on the edges of the honeycomb lattice automatically satisfy the branching rules explained above. Moreover, for every given set of string states on the honeycomb lattice that satisfy the branching rules, there are $N$ different dual states on the triangular lattice. The reason is transparent in Eq. (\ref{eq:dualmap1}): A universal shift of all spins on the triangular lattice by the same value leaves the dual honeycomb string state invariant. The modulo $N$ addition means the map is $N$ to 1. Recall that the string states satisfying the branching rules in the string--net models form only a subset of
the full Hilbert space. The Hamiltonian of a string--net model is chosen to ensure that those that satisfy the branching
rules are energetically favored. We thus at this point have a $N$ to 1 map between the
Hilbert space on the spin model, and the \emph{low lying} subspace of the Hilbert space of the string--net model.
The Hamiltonian of the SPT model is inherited from that of the string--net model built from the sum of magnetic flux operators
$B_p$ using this map. (The gauged version of
which involves only the insertion of a gauge field, to be explained in the next sub-section. ) The $N$ to 1 map therefore
immediately implies that a universal shift of all spin states at the vertices of the triangular lattice by the same value modulo
$N$ is guaranteed to commute with the Hamiltonian, which is insensitive to such a shift. The spin model therefore
enjoys a \emph{global} symmetry, which we will consider gauging in the next subsection.

Before we move on, let us remark that this duality of the honeycomb lattice with a triangular lattice
is well known in the context of topological quantum field theory (TQFT), where the triangular lattice
plays the role of a specific triangulation of the two dimensional surface\cite{Muegerl2010}. The construction
described above is therefore not special to the fusion algebra $\F_{\Z_N}$. It is
often emphasized that the orientation of the edges of the triangular lattice is chosen such that no closed
loop can be formed (see p.30 in Ref\cite{Baez2006}) We note however that our closed loops descends only
from a difference in the choice of convention. Our convention is such that the string states connecting each black (white) vertex on
the honeycomb lattice are all in-going (out-going), which led to a closed triangular loop as explained above.
If instead we choose a convention such that at each black (white) vertex there are two in-coming  (out-going) and one out-going (in-coming)
string state, the natural orientation of the dual triangular lattice would be such
that no closed loop can be formed. The triangulation is closely related to group cohomology \cite{Chen2011d}, which is not surprising
since as already discussed earlier, the structure emerges already in the classification of $6j$-symbols.

We now explain how gauging the symmetry that corresponds to the fusion algebra $\F_{\Z_N}$ on the
spin model provides a bijection between the complete Hilbert space of the string--net model
and that of the dual spin model.

\subsubsection{Gauging the dual spin model and counting of states}
Consider gauging the spin model on the triangular lattice $L_{\triangle}$. This is achieved by introducing to each edge connecting two neighbouring vertices, say, $a$ and $b$, a gauge field $\mu_{ab}$ that takes value in $\Z_N$. As a gauge theory, configurations related by gauge transformations are identified. The spins defined at the vertices transform as the fundamental matter in the model, whereas the gauge fields transform as the adjoint field. Each gauge transformation is a set $\{U_a\in\Z\vert\,\, a\in V(L_{\triangle})\}$, where $V(L_{\triangle})$ is the set of all vertices of the lattice. The transformation rule is given by
\be \label{eq:gaugetrans}
\begin{aligned}
\left(J_a+\tfrac{N-1}{2}\right) &\to& \left(J_a+\tfrac{N-1}{2}\right) + U_a\pmod{N}\,, \\
 \mu_{ab} &\to& \mu_{ab} + U_{a} - U_{b}\pmod{N}\,.
\end{aligned}
\ee

Note that in order to ensure well--defined modular arithmetic,
the combination $J_a + \tfrac{N-1}{2}$ appears in the definition of the gauge transformation, which
is always an integer for any spin $\tfrac{N-1}{2}$. Also, similar to the duality map
Eq. (\ref{eq:dualmap1}), the $(-)$ appearing in the gauge transformation for $\mu_{ab}$ is
assigned according to the chosen orientation on the link.
Now it is clear that there are as many gauge redundancies as the number of spin states
defined on the lattice sites. Suppose we fix a gauge such that
\be
J_a+\tfrac{N-1}{2}=0
\ee at
all sites $a$. Then it is
clear that each $\mu_{ab}$ is in 1--1 correspondence with the string degree of freedom defined on the
edge of the honeycomb lattice crossing the link $a-b$.
The map between states in the
string net model and those of the gauged spin model is now complete.

\subsubsection{The action of operators $B_p$}
The last salient component of the duality map is the action of the operators $B^s_p$
in the string--net model, where $p$ is a label of the hexagon
under consideration, and $s$ is one of the $N-1$ nonzero string types.
\footnote{For $s=0$ however,
$B^0_p$ is trivial i.e. equivalent to the identity operator.}
They form the building blocks of Hamiltonians that admit non--trivial
ground states supporting anyonic excitations.

Before we proceed, however, we need to understand the general map
between operators in the string--net model and those in the (gauged)
spin model.
As explained in previous paragraphs, every state on an
edge of the honeycomb lattice is determined by the spin states
on two neighbouring vertices of the dual edge in the triangular lattice, as given in
Eq. (\ref{eq:dualmap1}). This mapping, together with
the gauge transformation in Eq.(\ref{eq:gaugetrans}), suggests that
the natural spin operators acting on the spin degrees of freedom
would take the form
\be\label{eq:spinop}
\Sigma^z_\epsilon = e^{\ii \frac{2\pi}{N}(\eta^z_\epsilon + \frac{N-1}{2})},
\ee
where $\epsilon$ is a label of the spin site upon which
the spin resides. i.e., either
a vertex, or an edge on
the lattice; $\eta^{x,y,z}$ are
spin $\tfrac{N-1}{2}$ representations of  $SU(2)$ generators.

Therefore we have the following map between
spin operators in the string net model on
the honeycomb and those of the spin model on
the triangular lattice
\be \label{eq:mapop}
\Sigma^z\s{a} \to \Sigma^{z\,\dag}_{a}\Sigma^z_{ab}\Sigma^{z}_{b}.
\ee
Let us clarify that the operator $\Sigma^z_{ab}$ on the RHS above
acts on the gauge field $\mu_{ab}$ on the link connecting vertices $a$ and $b$.
The rest of the notation should be self--explanatory.

Let us also extend the cyclic shifting operator of string types defined in Eq. (\ref{eq:stringShiftOp}) to one that acts on the dual spins, $\Sigma^{\pm n}_{\epsilon}$ , which cyclically shifts the $J_z$ eigenvalue by $\pm n$-units (wrapping
around the bottom (top) $J_z$ states if $|J_z \pm n|> (N-1)/2$).
The only non-vanishing components are given by
\be
(\Sigma^{\pm n}_{\epsilon})_{\alpha \beta} = 1, \,\, \beta = \alpha \pm n \pmod{N}\,.
\ee
These operators share the same properties with those acting on the string states; they can thus also be constructed simply using the raising
and lowering operators of $SU(2)$. Gauge transformations defined by the set $\{U_a\}$ are achieved by acting with $\Sigma^{U_a}_{a}$ for fundamental matter,
and $\Sigma^{U_a-U_b}_{ab}$ on adjoint matter defined on the link
connecting vertices $a$ and $b$.

Gauge invariant operators on the triangular
lattice is built from
$\Sigma^{z\dag}_{a}\Sigma^z_{ab}\Sigma^{z}_{b}$,
precisely the combination appearing in the duality
map in Eq. (\ref{eq:mapop}).

Consider acting on a vertex $a$ on the triangular
lattice with $\Sigma^n_{a}$.
On the dual honeycomb lattice, we denote by $p_a$ the hexagon where the vertex $a$ is at  its center. The duality map Eq.(\ref{eq:dualmap1}) suggests
that this amounts to shifting the spins on
each edge $A$ of the 6 edges of $p_a$ by $\Sigma^{\sigma\s{a} n}_{s\s{a}}$,
where $\sigma\s{a} = \pm 1$. The sign depends on the assignment of
orientation of each link connecting two vertices cutting across the
edge $A$.
For instance, we can choose the convention in which $\sigma\s{a}$ takes ($+$) $-$ signs when the orientation of the edge $A$ on the honeycomb is (counter--) clockwise relative to the vertex $a$. We therefore have
\be
\prod_{A\in p_a} \Sigma^{{\sigma\s{a}}\,n}\s{a} = \Sigma^{n}_{a}
\ee

Taking into account the full duality map Eq. (\ref{eq:mapop})
after gauging the spin model on the triangular lattice, we can
see that the action of $\Sigma^{n}_{a}$ is
necessarily equivalent to that of  $\sideset{}{_{b\in V_a}}\prod \Sigma^{{\sigma_{ab}}n}_{ab}$,
where $V_a$ is the set of all the nearest neighbor vertices of $a$, and again
$\sigma_{ab} = \pm1$ according to the orientation of the link $ab$.
We therefore arrive at a gauge constraint:
\be
\prod_{b\in V_a} \Sigma^{{\sigma_{ab}}\,n}_{ab}= \Sigma^{n}_{a}
\ee

Having laid down the map between the $\Z_N$ string--net models on
the honeycomb lattice and gauged spin models on a triangular lattice
for general $N$, we would like to look into more details of the map
in some simple examples.
We need to understand the action of $B^s_p$
in the string net model after orientation has been fixed
through the split of the honeycomb lattice into two sub--latices
as explained above. Consider a vertex on a hexagon belonging to
sub--lattice $L\s{BL}$ labeled black. Like any other vertices,
three edges meet at this vertex,
two of which with string type $e_6$ and $e_1$ belongs to the
hexagon under consideration,
and a third leg $l_1$ is external to the hexagon (see Fig. \ref{fig:Duality}).
The action of $B_p$ at each vertex is such that it connects
the given state above to another state with string types
$e'_6, e'_1$ and $l'_1= l_1$ with an amplitude determined by
the $6j$-symbol $F^{l_1 e_6 e_1}_{s^* e'_1 e'_6}$. If the vertex
considered belongs to sub--lattice $L_{WH}$, the
corresponding amplitude is given by
$F^{l_1 e^*_6 e^*_1}_{s^* e'^*_1 e'^*_6}$.

While the amplitude can generally take non-trivial
values depending on the specific values of the $6j$-symbol,
it is important to note the relationship between
the primed string types and the unprimed ones.

The branching rules dictate that on a black vertex,
\be\label{eq:BpszN}
\begin{aligned}
e_1 - e'_1 + s =0 \pmod{N}\,, \\
e_6 - e'_6 - s =0 \pmod{N}\,.
\end{aligned}
\ee

On a white vertex a similar relation holds, with $s$ replaced
by $-s$. Note that the action of $B_p^s$ on neighboring vertices
on the hexagon acts consistently on their shared edge.
\subsection{$\F_{\Z_2}$ spin and string--net models}\label{subsec:revSNmodel}
The duality is best understood when $N=2$.
In this case, all the elements are self-dual. This
renders the assignment of orientations on the
string state entirely redundant. The choice
of orientation in the duality map in Eq.(\ref{eq:dualmap1})
also becomes redundant.
The spin operators in this case reduce to
\be
\Sigma^z_{\epsilon} = \sigma^z_\epsilon,\qquad \Sigma^{\pm 1}_\epsilon =\sigma^x_\epsilon,
\ee
where $\sigma^{z,x}_\epsilon$ are the usual Pauli matrices acting on
spin state $\epsilon$ residing either on a vertex or a link.

There are two $\F_{\Z_2}$ models, corresponding to two
different sets of $6j$ symbols. One is such that
all non-vanishing $6j$ symbols take value unity. In that case, the
$B_p^1$ operator and its spin model dual takes the form
\be
\prod_{A \in p_a} \sigma^x\s{a} \to \sigma^x_a\,,
\ee
where $a$ is the vertex in the triangular lattice located at the center
of the hexagon considered on the LHS of the map above.

The other $\F_{\Z_2}$ model, often referred to as the double--semion model,
whose $6j$ symbols can take different signs (see appendix),
the corresponding $B_{p_a}^1$ operator is given by
\be
B^1_{p_a}=\prod_{A \in p_a} \sigma^x\s{a}\prod_{B \in\ell(p_a)} \ii^{\frac{1-\sigma^z\s{b}}{2}}\to \sigma^x_a \prod_{b\in V_a} \ii^{\frac{1-\sigma^z_{a}\mu^z_{ab}\sigma^z_{b}}{2}}\,,
\ee
where $\ell(p)$ is the set of the six legs of the plaquette $p_a$ in the honeycomb lattice under consideration on the LHS.
\section{$\Z_3$ spin models and string--net models}\label{sec:z3SPmodel}
The $\Z_3$ is the next simplest example of the duality which exemplifies
some important features that are absent in $\Z_2$.
The most obvious deviation from the $\Z_2$ case is that not all string elements
are self-conjugate. This means the full machinery of orientation fixing
in the string net model via the split of the honeycomb into two sub--lattices $L\s{BL}$ and $L\s{WH}$,
and also the assignment of orientation to each link joining two vertices
in the dual triangular lattice becomes a necessity.
The allowed quantum dimensions and $6j$ symbols are given in the appendix.

The model retains some simplicity.
We observe that the $J^z$ matrix satisfies $(J^z)^3 = J^z$.
Therefore the spin operator $\Sigma^z_\epsilon$ defined in Eq.(\ref{eq:spinop})
simplifies to
\be
\Sigma^z_{\epsilon} = e^{\ii \frac{2\pi}{3}}\left[ P_0 + P_{\pm1} \cos(\tfrac{2\pi}{3}) + \ii J^z\sin(\tfrac{2\pi}{3})\right]_\epsilon\,,
\ee
where $P_{0}$ and $P_{\pm 1}$ are respectively the projectors to the subspaces of states with $J^z = 0$ and $J^z\in\{\pm1\}$.

In the case where all $6j$ symbols equal unity, $B^s_p$ is given by
\be
B^s_p = \prod_{A \in p_{-}} \Sigma^{-s}\s{a}\prod_{B \in p_+} \Sigma^{+s}\s{b} \to \Sigma^s_{a}
\ee
where the edges of the hexagonal plaquette is split into two groups $p_-$ and $p_+$ according to our choice of orientation.
See Eq. (\ref{eq:BpszN}).

%\subsection{spin model and the duality}\label{subsec:SPdual}

%\section{$\Z_n$ Lattice spin models}\label{sec:znSPmodel}
%Zn generalization, perhaps with less details than the previous two sections but mention/list the generalization of the necessary components namely:
%\begin{itemize}
%\item $6j$ symbols: nonvanishing ones all equal to 1. II think we can prove it!
%\item $B_p$: probably not too  much detail. perhaps mention in words
%\item The spin model.
%\end{itemize}

\section{Discussions and Outlook}\label{sec:outlook}
In this paper, we hope to achieve three different goals. First, we explored in
some detail the solutions of $6j$ symbols to the pentagon relations particularly
for fusion rules given by the $\Z_N$ group. Given that the study of topological
order is a relatively new subject, we believe it is a valuable exercise to supply more explicit
details of simple examples that forms the basis of intuition and expectation
of the physics of these classes of models. Second, by revisiting the basic features
of the $6j$ symbols, we hope to make more explicit and physically intuitive how
their classification is related to the Mathematics of group cohomology, at least
in the case where the fusion algebra forms a group. While we later realize that this
is a well known fact in tensor category theory, we believe that such a discussion
that anchors in a more physical setting is valuable. Nevertheless, we would like to emphasize
that such a connection with group cohomology has so far been achieved when
the fusion algebra does coincide a group. In the more general situation where a fusion algebra can only be viewed as
a \emph{hypergroup} and where multiplicites $\mathcal{N}^k_{ij}>1$, a
general classification of the possible $6j$ symbols is lacking, and is an open
problem in the theory of fusion categories.

Third, we pick up on the trail
first opened up in \cite{Levin2012, Burnell2011} to supply further evidence of the connection
between SPT phases and LRE phases. We construct a general map between an SPT
phase whose global symmetry $\Z_N$ is \emph{gauged} and $\F_{\Z_N} $ string--net
models, generalizing the special case explored in Ref\cite{Gu2012} of $N=2$.
It is not hard to see that our map between states in the SPT phase and the string--net
model applies also to other (non) Abelian finite groups. It serves as a confirmation
that indeed the appearance of $H^3(G,U(1))$ in both the classification of these
SPT phases \cite{Chen2011d} and that of the string net models is by no means an accident.

There are many interesting and important problems that should be addressed. One immediate
question of interest is to understand the quasi-particle excitations
of these $\F_{\Z_N} $ string--net models, and investigate, using the map we have
constructed, their relationship with edge excitations of SPT phases. In this paper,
as an initial attempt we have focused on $6j$ symbols that respect the full
tetrahedral symmetry, and as discussed in the text, this is a very restrictive choice which
excludes possible solutions of the pentagon relations. It would be important to look for
more general models that break the tetrahedral symmetries.
More fundamentally, it would be of great importance to have a more general classification
of these string--net models when the fusion algebra involved does not form
a group. We hope to return to some of these questions in the near future.

\begin{acknowledgments}
We thank
Oliver Buerschaper, Zhengcheng Gu, Yuting Hu, Xiaogang Wen, and  Yong-shi Wu for helpful and encouraging discussions! YW appreciate the Perimeter Institute for hospitality during his visit when part of this work was done.
Research at Perimeter Institute is supported by the Government of Canada through Industry Canada and by the Province of Ontario through the Ministry of Research \& Innovation. YW is partially supported by \lq Open Research Center\rq~Project for Private Universities: matching fund subsidy from the Ministry of Education, Culture, Sports, Science and Technology, Japan (MEXT).
The authors also thank Department of Physics and Center for Field Theory and Particle Physics, Fudan University (Shanghai, China) for warm hospitality during a mini workshop on "Exactly Solvable Discrete Models for Topological Matter" in summer 2012, where the completion of the manuscript was done.
\end{acknowledgments}
\appendix
\section{$6j$ Symbols of $\F_{\Z_N}$ String--net models}\label{appx:6jSymbols}
In this appendix, we compute the $\F_{\Z_N}$ $6j$  symbols explicitly in general, by following the results in Section \ref{subsec:zn6jQDim}. Then we shall present the cases where $N=3,4, \text{ and } 6$ as examples.

For the sake of constructing and understanding the magnetic flux operators $B_p^s$ discussed in Section, we would like to change the variables in the canonical forms of $G$-- and $F$--symbols, defined in Eqs. (\ref{eq:2Gunity}) and (\ref{eq:2Funity}), which we now rewrite as follows.
\begin{align}
G^{l^{}\s{A}\, e^*\s{A}\, e^{}\s[-1]{A}}_{s^* (e\s[-1]{A}+s) (e\s{A}+s)^*}G^{l^*\s{A}\, e^{}\s{A}\, e^*\s[-1]{A}}_{s(e\s[-1]{A}+s)^* (e\s{A}+s)}d_{l\s{A}}d_s=1\label{eq:2GunityNew}\\
\left|F^{l^{}\s{A}\, e^*\s{A}\, e^{}\s[-1]{A}}_{s^* (e\s[-1]{A}+s) (e\s{A}+s)^*}\right|^2=\left|G^{l^{}\s{A}\, e^*\s{A}\, e^{}\s[-1]{A}}_{s^* (e\s[-1]{A}+s) (e\s{A}+s)^*}\right|^2 \equiv 1\label{eq:2FunityNew}
\end{align}
where the substitutions $n^*=l\s{A}$, $i=s^*$, $p=e^*\s{A}$, and $l+i=e_{A-1}\pmod{N}$, with the index $A$ running over the vertices of a hexagon, are made in Eqs. (\ref{eq:2Gunity}) and (\ref{eq:FsymbNorm1}). The first equality in Eq. (\ref{eq:2FunityNew}) follows simply
from the fact that all $v_i$ also have unit norm.
The canonical $F$--symbol with the new indices in Eq. (\ref{eq:2FunityNew}) plays a key role in manifesting the action of the magnetic flux operators. We now compute the $\F_{\Z_N}$ $F$--symbols in their canonical form.

Equation (\ref{eq:2FunityNew}) shows that a generic $F$--symbol is a complex number whose norm is one. But some $F$-symbols are constrained to be real, but can be either $\pm1$, while some others are fixed uniquely to be one, as is to be shown in the following.

According to Eq. (\ref{eq:znFtetsign}) and the discussion beneath the equation, if an $F$--symbol is real or pure imaginary, it can be transformed into its conjugate by   the tetrahedral  transformation. This fact indicates that Eqs. (\ref{eq:2GunityNew}) and (\ref{eq:2FunityNew}) are handy for determining the values of the $\F_{\Z_N}$ $6j$ symbols. We shall rely on Eq. (\ref{eq:2GunityNew}) more because the $G$--symbols are manifestly symmetric and then obtain the $F$-symbols from the $G$--symbols by Eq. (\ref{eq:F2G6j}). We shall compute the canonical $F$--symbols only for two reasons: First, the non--canonical ones can be obtained from the canonical ones by the tetrahedral transformation; Second, in Section \ref{subsec:ZnStringOp}, one can see that the $F$--symbols that appear in the magnetic flux operators and hence in the Hamiltonians of $\F_{\Z_N}$ string--net models are always in the canonical form.

Let $s=0$ in the canonical $G$-symbols. Then we have
\be\label{eq:znGs=0}
G^{l\s{A}\, e^*\s{A}\, e^{}\s[-1]{A}}_{0\, e^{}\s[-1]{A} e^*\s{A}}=G^{l^*\s{A}\, e^{}\s{A}\, e^*\s[-1]{A}}_{0\,e^*\s[-1]{A}e^{}\s{A}}=G^{\, e^{}\s[-1]{A}e^*\s{A}l^{}\s{A}\, }_{e\s{A}\,e\s[-1]{A}\, 0}=\frac{1}{v_{e\s{A}}v_{e\s[-1]{A}}}\,,
\ee
where $l\s{A}=e\s{A}-e\s[-1]{A}\pmod{N}$ is understood. Clearly, Eq. (\ref{eq:2GunityNew}) is trivially satisfied in this case. The corresponding canonical $F$--symbols is readily
\be\label{eq:znFs=0}
F^{l^{}\s{A}\, e^*\s{A}\, e^{}\s[-1]{A}}_{0\, e^{}\s[-1]{A} e^*\s{A}}=G^{l^{}\s{A}\, e^*\s{A}\, e^{}\s[-1]{A}}_{0\, e^{}\s[-1]{A} e^*\s{A}}v_{e^*\s{A}}v_{e^{}\s[-1]{A}} =\frac{v_{e^*\s{A}}v_{e^{}\s[-1]{A}}}{v_{e\s{A}}v_{e\s[-1]{A}}}\equiv 1,\,
\ee
where Eq. (\ref{eq:F2G6j}) and 
\be\label{eq:vv*}
v_{e^*\s{A}}=v_{e\s{A}}
\ee 
are applied. Let us note here that very generally $d_{e^*\s{A}}=d_{e\s{A}}$
since any loop of a given orientation can be smoothly deformed to its opposite orientation on a sphere, and since 
we define $v_{e\s{A}}= \sqrt{d_{e\s{A}}}$, which suffers some redundancy, one can impose the convention (\ref{eq:vv*}). %JH

This is an important identity, as it shows that the magnetic flux operator $B_p^{s=0}\equiv 1$ in Section \ref{subsec:ZnStringOp}.

Instead of setting $s$ to be zero, we can also set either of $l\s{a}$, $e\s{a}$, or $e\s[-1]{A}$ be zero.

First, let $l\s{a}=0$, which then requires that $e\s{a}=e\s[-1]{a}$ and renders Eq. (\ref{eq:2GunityNew}) as
\[
G^{0\, e^*\s{A}\, e^{}\s[-1]{A}}_{s^* (e\s[-1]{A}+s) (e\s{A}+s)^*}G^{0\,e^{}\s{A}\, e^*\s[-1]{A}}_{s(e\s[-1]{A}+s)^* (e\s{A}+s)}d_s=1,
\]
where we  leave $e\s{a}$ and $e\s[-1]{a}$ unidentified to emphasize their ordering. It is easy to show by the tetrahedra transformation and the equality $e\s{a}=e\s[-1]{a}$ that the two $G$--symbols in the equation above are equal.
This equation is actually trivially satisfied, as can be checked straightforwardly. Hence, we obtain
\be\label{eq:znGeA=0}
G^{0\, e^*\s{A}\, e^{}\s[-1]{A}}_{s^* (e\s[-1]{A}+s) (e\s{A}+s)^*} =\frac{1}{v_{e\s{a}}v_{e\s{a}+s}}\;,
\ee
which is real if $d_s=1$ and imaginary if $d_s=-1$, as $v_{e\s{a}}v_{e\s{a}+s}=\pm\sqrt{d_{s}}$. The corresponding canonical $F$--symbol is then
\be\label{eq:znFeA=0}
F^{0\, e^*\s{A}\, e^{}\s[-1]{A}}_{s^* (e\s[-1]{A}+s) (e\s{A}+s)^*} =G^{0\, e^*\s{A}\, e^{}\s[-1]{A}}_{s^* (e\s[-1]{A}+s) (e\s{A}+s)^*}v_{e\s{a}}v_{e\s{a}+s}\equiv 1
\ee

Second, take $l\s[-1]{a}=0$. As such, Eq. (\ref{eq:2GunityNew}) becomes
\[
G^{l^{}\s{A}\, e^*\s{A}\, 0}_{s^* s (e\s{A}+s)^*}G^{l^*\s{A}\, e^{}\s{A}\, 0}_{ss^* (e\s{A}+s)}d_{l\s{A}}d_s=1,
\]
where $e\s{a}=l\s{a}$ is understood. This equation is trivially satisfied, as can be verified by the tetrahedral symmetry---that shows the two $G$--symbols are equal---and Eq. (\ref{eq:G6jRelNor}). Hence, by Eq. (\ref{eq:G6jRelNor}) we have
\be\label{eq:znGlA-1=0}
G^{l^{}\s{A}\, e^*\s{A}\, 0}_{s^* s (e\s{A}+s)^*}=\frac{1}{v_{l\s{a}}v_s}\;,
\ee
which takes value in $\{\pm 1,\pm \ii\}$, depending on $v_{l\s{a}}$, and $v_s$. But the corresponding canonical $F$--symbol is clearly real:
\be\label{eq:znFlA-1=0}
F^{l^{}\s{A}\, e^*\s{A}\, 0}_{s^* s (e\s{A}+s)^*} =
\frac{v_{e\s{a}+s}}{v_{l\s{a}}v_s}\biggr\vert_{e\s{a}=l\s{a}} =\pm 1.
\ee

By symmetry, the canonical $G$--symbol in the case where $l\s{a}=0$, equals that in Eq. (\ref{eq:znGlA-1=0}), namely
\be\label{eq:znGlA=0}
G^{l^{}\s{A}\, 0\, e\s[-1]{A}}_{s^*  (e\s[-1]{A}+s) s^*}=\frac{1}{v_{l\s{a}}v_s}\;;
\ee
however, the corresponding canonical $F$--symbol turns out to be just unity:
\be\label{eq:znFlA=0}
F^{l^{}\s{A}\, 0\, e\s[-1]{A}}_{s^*  (e\s[-1]{A}+s) s^*} = \frac{v_{e\s[-1]{a}}v_s}{v_{l\s{a}}v_s} \biggr\vert_{l^*\s{a}=e\s[-1]{a}}\equiv 1.
\ee
This result can also be obtained from Eq. (\ref{eq:znFlA-1=0}) by the tetrahedral symmetry and renaming of repeated indices. Yet we list Eq. (\ref{eq:znFlA=0}) as a stand-alone result again for our convenience of constructing the magnetic flux operators.

We now check the cases where none of the indices of the canonical $G$--symbols
in Eq. (\ref{eq:2GunityNew}) is fixed to be zero. In particular we look for the cases where the second $G$-symbols in Eq. (\ref{eq:2GunityNew}) can be turned into the first $G$-symbol in the equation by the tetrahedral transformations, such that the numerical values of the $G$--symbol can be determined.

First, consider that $l\s{A}=s^*\neq0$, it follows from the branching rules that $s=e\s[-1]{A}-e\s{A}$. Hence, also by Eq. (\ref{eq:qDimSquare1}), Eq. (\ref{eq:2GunityNew}) becomes
\[
G^{s^*\,e^*\s{A}\, (e^{}\s{A}+s)}_{s^* (e\s{A}+2s) (e\s{A}+s)^*}G^{s\, e\s{A}\,  (e\s{A}+s)^*}_{s(e^{}\s{A}+2s)^* (e\s{A}+s)}=1,
\]
where the $G^{s^*\,e^*\s{A}\, (e^{}\s{A}+s)}_{s^* (e\s{A}+2s) (e\s{A}+s)^*}=G^{s\,e\s{A}\,  (e\s{A}+s)^*}_{s\,(e^{}\s{A}+2s)^* (e\s{A}+s)}$ can be easily verified by using Eq. (\ref{eq:G6jRelTet}).
Thus the equation yields
\be\label{eq:znGs*s*}
G^{s^*\,e^*\s{A}\, (e^{}\s{A}+s)}_{s^* (e\s{A}+2s) (e\s{A}+s)^*} =\pm 1\,,
\ee
giving rise to the corresponding canonical $F$--symbols as
\be\label{eq:znFs*s*}
F^{s^*\,e^*\s{A}\, (e^{}\s{A}+s)}_{s^* (e\s{A}+2s) (e\s{A}+s)^*} =\pm v^2_{e\s{A}+s}=\pm d_{e\s{A}+s}\,=\pm 1,
\ee
which is certainly real but depends on the choice of $d_{l\s{A}+s}$.

Second, let $l\s{A}=s\neq 0$, which, together with Eq. (\ref{eq:qDimSquare1}), implies that
\[
G^{s\,e^*\s{A}\, (e^{}\s{A}-s)}_{s^* \,e\s{A} (e\s{A}+s)^*}G^{s^* e\s{A}\,  (e\s{A}-s)^*}_{s\,e^*\s{A} (e\s{A}+s)}=1,
\]
where the equality $G^{s\,e^*\s{A}\, (e^{}\s{A}-s)}_{s^* \,e\s{A} (e\s{A}+s)^*}=G^{s^* e\s{A}\,  (e\s{A}-s)^*}_{s\,e^*\s{A} (e\s{A}+s)}$ is guaranteed by Eq. (\ref{eq:G6jRelTet}), which means
\be\label{eq:znGss*}
G^{s \,e^*\s{A}\, (e^{}\s{A}-s)}_{s^* e\s{A} (e\s{A}+s)^*} = \pm 1.
\ee
The corresponding $F$-symbol is then
\be\label{eq:znFss*}
F^{s \,e^*\s{A}\, (e^{}\s{A}-s)}_{s^* e\s{A} (e\s{A}+s)^*} = \pm v_{e^{}\s{A}-s} \,v_{e\s{A}+s}\;,
\ee
which is also real because
\[
v_{e^{}\s{A}-s} \,v_{e\s{A}+s}
=\pm \sqrt{d_{e^{}\s{A}-s} \,d_{e\s{A}+s}}
=\pm \sqrt{d_{2e\s{a}}}
=\pm 1\;
\]
Note again that $s\neq 0$ is assumed in Eqs. (\ref{eq:znFs*s*}) and (\ref{eq:znFss*}); if we demanded that $s=0$, the $F$--symbols reduce to that in Eq. (\ref{eq:znFeA=0}) and is thus single-valued.

By the tetrahedral symmetry and possible renaming of repeated indices, we have summarized all the canonical $F$--symbols that are either strictly unity or admit either solutions $\pm1$, and one notes that they happen to be all real. That said, complex $F$--symbols can arise when their canonical forms do not constrain $s$, $l\s{a}$, $e\s{a}$, or $e\s[-1]{a}$. Moreover, complex $F$--symbols do not occur for $N\leq 5$, as the cases discussed above are all that can appear for $N\leq 5$. In other words, every $F$--symbol of $N\leq 5$ is real. Let us prove this statement.

Recall that we have been dealing with canonical $6j$ symbols of $\F_{\Z_N}$, from which all other non-vanishing $6j$ symbols can be obtained by the tetrahedral transformations. This implies that the various cases of the canonical $G$--symbol , $G^{l^{}\s{A}\, e^*\s{A}\, e^{}\s[-1]{A}}_{s^* (e\s[-1]{A}+s) (e\s{A}+s)^*}$, discussed so far in this appendix
can be grouped and phrased as the two classes as follows.
\begin{enumerate}
\item Any index of the $G$--symbol
is zero.
\item The two indices in any column are either identical or conjugate of each other.
\end{enumerate}
We want to show that any $G$--symbol of $\F_{\Z_N}$ for $N\leq 5$ must fall into either of the two classes. It is however more convenient to prove that a $G$--symbol that violates the criteria of both classes can arise only for $N\geq 6$.

The branching rule $\delta_{l^{}\s{A}\, e^*\s{A}\, e^{}\s[-1]{a}}=1$ enables us to rewrite the $G$--symbol explicitly as $G^{l^{}\s{A}\, e^*\s{A}\, (e^{}\s{A}+l^*\s{a})}_{s^* (e\s{A}+l^*\s{a}+s) (e\s{A}+s)^*}$ and express the conditions leading to the $G$--symbols not belonging to any of the two classes enlisted above as the following six independent and complete inequalities.
\bse\label{ieq:s}
\begin{eqnarray}
&s &\neq 0\label{ieq:s1}\\
&s &\neq l\s{a}\label{ieq:s2}\\
&s &\neq l^*\s{a}\label{ieq:s3}\\
&s &\neq l\s{a}+e^*\s{a}\label{ieq:s4}\\
&s &\neq e^*\s{a}\label{ieq:s5}\\
&s &\neq 2e^*\s{a}+l\s{a}\label{ieq:s6}\;,
\end{eqnarray}
\ese
where $l\s{a}\neq 0$, $e\s{a}\neq 0$, and $l\s{a}\neq e\s{a}$ are assumed. Inequalities (\ref{ieq:s}) are independent because they cannot derive each other. The point is then to demonstrate that a solution of the string type $s$ that meets all the six inequalities in (\ref{ieq:s}) given $l\s{a}$ and $e\s{a}$ exists only for $N\geq 6$. This is in fact rather obvious, as in cases where $N\leq 5$, there are at most five string types including zero, which are certainly insufficient to give a solution to the six inequalities (\ref{ieq:s1}) through (\ref{ieq:s6}).

On the other hand, if $N=6$, there are six string types all told, which can solve the six inequalities. Here is an example: Let $l\s{a}=1$ and $e^*\s{a}=1$, the immediate solution to inequalities (\ref{ieq:s}) is $s=4$, which yields the canonical $G$--symbol $G^{114}_{223}$ being the first $G$--symbol in Eq. (\ref{eq:2GunityNew}), where the second $G$--symbol is accordingly $G^{552}_{443}$. Clearly, $G^{114}_{223}$ and $G^{552}_{443}$ can never be identified by the tetrahedral transformations, which renders the $2G$ relation Eq. (\ref{eq:2GunityNew}) incapable of dictating the values of the two $G$--symbols. As a consequence, the corresponding $F$--symbol $F^{114}_{223}$ is complex with norm one by Eq. (\ref{eq:2FunityNew}). This conclusion actually applies to any $F$--symbol whose indices satisfy all the inequalities in (\ref{ieq:s}).

As $N$ grows, inequalities (\ref{ieq:s}) have more solutions and thus giving rise to more complex $F$--symbols.
We highlight this result as a theorem.
\begin{theorem}\label{theo:znFsymb}
In $\F_{\Z_N}$ string--net models with $N\geq 6$, there are $F$--symbols that are complex with $|F|=1$. In cases where $N\leq 5$, all $F$--symbols are either strictly one, or allowing both $\{\pm 1\}$.
\end{theorem}
According to the discourse in this section, we may write a generic $\Z_N$ canonical $F$--symbol as
\be\label{eq:znFasPhase}
F^{l^{}\s{A}\, e^*\s{A}\, e^{}\s[-1]{A}}_{s^* (e\s[-1]{A}+s) (e\s{A}+s)^*}=\e^{\ii \alpha\s{n}(s,l\s{a},e\s{a})},
\ee
where $\alpha\s{n}(s,l\s{a},e\s{a})$ is a function of $s$, $l\s{a}$, and $e\s{a}$ with subscript $N$. Therefore, the results in this section can be summarized as the following equation.
\be\label{eq:znFangle}
\begin{aligned}
\alpha\s{n}(0,l\s{a},e\s{a}) &=\alpha\s{n}(s,0,e\s{a})\\
 &=\alpha\s{n}(s,l\s{a},0)=\alpha\s{n}(s,s,s)\equiv 0\;,\\
\alpha\s{n}(s,l\s{a},l\s{a})&\in\{0,\pi\}\;,\\
\alpha\s{n}(s,s^*,e\s{a})&\in\{0,\pi\}\;,\\
\alpha\s{n}(s,s,e\s{a})&\in\{0,\pi\}\;,\\
\end{aligned}
\ee
which are true for all $N$ and are all that can happen in cases where $N\leq 5$. When $N\geq 6$, we are lack of further constraints on the function $\alpha\s{n}(s,l\s{a},e\s{a})$ and shall leave it as a free parameter of the model.

As aforementioned, the fusion algebra $\F_{\Z_N}$ can be shared by some other (quantum) groups; therefore, each set of solutions to Eqs. (\ref{eq:F6jRel}) shown  in this section may be the $F$--symbols of one of these groups. Since notationally we based the algebra $\F_{\Z_N}$ on $\Z_N$, one may want to obtain the $Z_N$ $F$--symbols more explicitly, which are derived from another point of view in the following subsection.
\subsection{$\Z_N$ $F$--symbols from $3j$ symbols}
In this appendix, we show that the $F$--symbols of the group $\Z_N$ are all unity by deriving them from the $3j$ symbols of $\Z_N$. For the sake of generality, we consider the irreducible, projective  representations of $\Z_N$, which are one--dimensional and can be labeled by integers $0,\ 1,\ 2,\ \dots,\ N-1$. Let $j=\e^{\ii\vartheta_j}$ be the $j$--th element in the projective representation labeled by $\vartheta$, such that $\vartheta_j$ is the phase associated with $j$. Since the $Z_N$ fusion rule is  $j\otimes k=j+k\pmod{N}$, we would need to equate $\e^{\ii\vartheta_j} \e^{\ii\vartheta_k}=\e^{\ii(\vartheta_j +\vartheta_k)}$ and $\e^{\ii\vartheta_{j+k}}$; hence, the $3j$ symbols of $\Z_N$ is
\be
\e^{i(\vartheta_j+\vartheta_k-\vartheta_{j+k})},
\ee
which indeed complies with the fusion algebra:
\[
\e^{\ii\vartheta_j} \e^{\ii\vartheta_k}=\e^{\ii\vartheta_{j+k}} \e^{i(\vartheta_j+\vartheta_k-\vartheta_{j+k})}.
\]
Now we can construct the $6j$ symbols, i.e., the $F$--symbols from the $3j$ symbols by the associativity of tensor products, namely,
\[
(i\otimes j)\otimes k=l=
i\otimes (j\otimes k).
\]
Let $i\otimes j=m$ and $j\otimes k=n$. Then in terms of the projective representations and $3j$ symbols, the LHS and the RHS of the above equation are respectively
\[
\e^{\ii(\vartheta_i+\vartheta_j-\vartheta_m)} \e^{\ii\vartheta_l} \e^{\ii(\vartheta_m+\vartheta_k-\vartheta_l)}     =          \e^{\ii(\vartheta_i+\vartheta_j+\vartheta_k-\vartheta_l)}    \e^{\ii\vartheta_l}
\]
and
\[
\e^{\ii(\vartheta_j+\vartheta_k-\vartheta_n)} \e^{\ii\vartheta_l} \e^{\ii(\vartheta_i+\vartheta_n-\vartheta_l)}     =          \e^{\ii(\vartheta_i+\vartheta_j+\vartheta_k-\vartheta_l)}    \e^{\ii\vartheta_l},
\]
which are already equal. Therefore, we have for $\Z_N$,
\be
F\equiv 1\,.
\ee

\subsection{$\F_{\Z_3}$ $6j$ symbols}\label{appx:6jZ3}
Since we would like this paper to serve as a concise review of and reference to string-net models, in this $\Z_3$ example, we shall compute the $6j$ symbols as if we did not have the general results of $\Z_N$ obtained in the previous section but do so by naively following the first principles of computing $6j$ symbols. In subsequent sections where $\Z_4$, $\Z_6$ are studied, we shall directly list the non-vanishing $G$-- and $F$--symbols acquired by applying the general results in the previous section.

In the case of $\Z_3$, the branching rules are $\{0,0,0\}$, $\{0,1,2\}$, $\{1,1,1\}$, and $\{2,2,2\}$. We are looking only for the non--vanishing $F$--symbols. We study this in two cases, where the string type $m$ of the horizontal edge on the LHS of Eq. (\ref{eq:SNWaveFuncCross}) is respectively $0$ and $1$. There is in fact a third case, i.e., $m=2$; however, since $2=1^*$, the corresponding $F$--symbols can be obtained from those in the case $m=1$ by either complex conjugation or tetrahedral symmetry.

\textbf{Case 1}: $m=0$. The crossing symmetry now reads
\be\label{eq:z3Fsymb-m0}
\BLvert
\Hgraph{j}{0}
\Brangle=F^{j_1j_20}_{j_3j_4n}\BLvert
\Xgraph{j}{n}
\Brangle.
\ee
In view of the branching rules, $j_1$, $j_2$, $j_3$, and $j_4$ can take only these values: $(1,2,1,2)$, $(1,2,2,1)$, $(2,1,1,2)$, and $(2,1,2,1)$. The latter two are obviously the conjugate of the former two, we thus need only to study the former two options. Hence, accordingly, the string type on the vertical edge on the RHS of Eq. (\ref{eq:z3Fsymb-m0}) must be $m=0$ for $(1,2,1,2)$ and $m=1$ for $(1,2,2,1)$, which give rise to non-vanishing $F$--symbols respectively, namely $F^{120}_{120}$ and $F^{120}_{211}$, which are clearly physically distinct. The corresponding $G$--symbols, $G^{120}_{120}$ and $G^{120}_{211}$ are non-vanishing as well. Because the tetrahedral symmetry is manifest in the $G$--symbols, these two $G$--symbols immediately lead to the other non-vanishing $G$--symbols in this case as follows, according to Eq. (\ref{eq:G6jRelTet}).
\be\label{eq:z3Gsymb-m0}
\begin{aligned}
&G^{120}_{120}=G^{210}_{210}=G^{021}_{012}=G^{201}_{101}=G^{012}_{021}=G^{102}_{202}=\frac{1}{v_1^2}\\
&G^{120}_{211}=G^{210}_{122}=G^{021}_{221}=G^{201}_{222}=G^{012}_{112}=G^{102}_{111}=\frac{1}{v_1^2}\\
&=G^{222}_{022}=G^{222}_{201}=G^{111}_{011}=G^{111}_{201}=G^{111}_{220}=G^{222}_{110},
\end{aligned}
\ee
where the $\frac{1}{v_1^2}$ is due to Eq. (\ref{eq:G6jRelNor}).

\textbf{Case 2}: $m=1$. The crossing symmetry in this case is \be\label{eq:z3Fsymb-m1}
\BLvert
\Hgraph{j}{1}
\Brangle=F^{j_1j_20}_{j_3j_4n}\BLvert
\Xgraph{j}{n}
\Brangle,
\ee
which, together with the branching rules, results in nine non-vanishing $F$--symbols that directly appear in this equation, i.e., $F^{111}_{220}$, $F^{111}_{102}$, $F^{111}_{011}$, $F^{201}_{222}$, $F^{201}_{101}$, $F^{201}_{010}$, $F^{021}_{221}$, $F^{021}_{100}$, and $F^{021}_{012}$. Obviously, among these nine $F$--symbols, only one -- $F^{201}_{010}$ -- has its $G$--symbol distinct from those listed in Eq. (\ref{eq:z3Gsymb-m0}). Hence, the remaining non-vanishing $G$--symbols are
\be\label{eq:z3Gsymb-m1}
\begin{aligned}
G^{201}_{010}&=G^{021}_{100}=G^{102}_{020}=G^{120}_{002}\\
&=G^{210}_{001}=G^{012}_{200}=G^{000}_{121}=G^{000}_{212}=\frac{1}{v_1},
\end{aligned}
\ee
where the last equality is due to Eq. (\ref{eq:G6jRelNor}).

There is but one more nonzero $G$--symbol by definition: $G^{000}_{000}\equiv 1$. The value of $v_1$, the square root of the quantum dimension, can now be determined by the $2G$ relation Eq. (\ref{eq:2Grel}) as follows. Let us set $i=q=0$ in Eq. (\ref{eq:2Grel}), which then requires $m=l^*$ and $k=p$ by the branching rules. Choosing $l=k=1$, Eq. (\ref{eq:2Grel}) becomes
\[
d_2\left(G^{210}_{122}\right)^2=\frac{1}{v_1^2}=1\Longrightarrow v_1=\pm 1,\ d_1=1,
\]
where use of Eq. (\ref{eq:z3Gsymb-m1}) is made in the first equality. Equations (\ref{eq:z3Gsymb-m0}), (\ref{eq:z3Gsymb-m1}), and (\ref{eq:F2G6j}) yield the following non-vanishing $F$--symbols, which are grouped by their values instead of the tetrahedral symmetry.
\bse\label{eq:z3FsymbNon0}
\begin{eqnarray}
&F^{000}_{000} &=F^{120}_{120}=F^{210}_{210}=F^{021}_{012}=F^{201}_{101}\\
& &=F^{012}_{021}=F^{102}_{202}=F^{021}_{221}=F^{201}_{222}=F^{012}_{112}\nonumber\\
& &=F^{102}_{111}=F^{222}_{022}=F^{222}_{201}=F^{111}_{011}=F^{111}_{201}=1\nonumber\\
&F^{120}_{211} &=F^{210}_{122}=F^{111}_{220}=F^{222}_{110}=\frac{1}{v_1}=\pm1.
\end{eqnarray}
\ese
All the $\F_{\Z_3}$ $F$--symbols are thus real, which is why in this case an $F$--symbol and its conjugate are related by the tetrahedral symmetry. The choice that $F\equiv1$  gives the $F$--symbols of  the group $\Z_3$.
\subsection{$\F_{\Z_4}$ $6j$ symbols}\label{appx:6jZ4}
The $\F_{\Z_4}$ string--net model bears four string types, i.e., $0$, $1$, $2$, and $3$, and the branching rules $\{0,0,0\}$, $\{0,1,3\}$, $\{0,2,2\}$, $\{1,1,2\}$, and $\{2,3,3\}$.

We first collect all the non-vanishing $G$--symbols, grouped by the tetrahedral symmetry.
\begin{eqnarray}\label{eq:z4GsymbNon0}
& G^{202}_{202} &=G^{022}_{022}=G^{220}_{220}=\frac{1}{v_2^2}\nonumber\\
& G^{022}_{200} &=G^{202}_{020}=G^{220}_{002}=G^{000}_{222}=\frac{1}{v_2}\nonumber\\
& G^{130}_{130} &=G^{013}_{013}=G^{301}_{101}=G^{031}_{013}=G^{103}_{303}=G^{310}_{310}=\frac{1}{v_1^2}\nonumber\\
& G^{031}_{100} &=G^{103}_{030}=G^{310}_{001}=G^{000}_{313}\nonumber\\
&               &=G^{301}_{010}=G^{130}_{003}=G^{013}_{300}=G^{000}_{131}=\frac{1}{v_1}\\
& G^{310}_{132} &=G^{031}_{231}=G^{103}_{121}=G^{130}_{312}\nonumber\\
&               &=G^{013}_{213}=G^{301}_{323}=G^{323}_{301}=G^{332}_{110}\nonumber\\
&               &=G^{233}_{033}=G^{112}_{330}=G^{211}_{011}=G^{121}_{103}=\frac{1}{v_1^2}\nonumber\\
& G^{112}_{203} &=G^{211}_{320}=G^{121}_{012}=G^{202}_{111}=G^{220}_{133}\nonumber\\
&               &=G^{022}_{331}=G^{332}_{023}=G^{233}_{302}=G^{323}_{210}=G^{310}_{223}\nonumber\\
&               &=G^{103}_{212}=G^{031}_{322}=G^{323}_{032}=G^{233}_{120}=G^{332}_{201}\nonumber\\
&               &=G^{202}_{333}=G^{220}_{311}=G^{022}_{113}=G^{121}_{230}=G^{112}_{021}\nonumber\\
&               &=G^{211}_{102}=G^{130}_{221}=G^{013}_{122}=G^{301}_{232}=\frac{1}{v_1 v_2}\nonumber\\
& G^{121}_{321} &=G^{112}_{112}=G^{211}_{233}=G^{233}_{211}=G^{332}_{332}=G^{323}_{123}=\pm\frac{1}{d_2},\nonumber
\end{eqnarray}
where
\be\label{eq:z4quantDim}
\begin{cases}
v_1=v_3=\pm 1,\ \pm\ii & \Rightarrow\quad d_1=d_3=\pm 1,\\[0.5em]
v_2=\pm1 & \Rightarrow\quad d_2\equiv 1.
\end{cases}
\ee
 We gather the unity $F$--symbols in the equations below.
\begin{eqnarray}\label{eq:z4FsymbNoSign}
& F^{202}_{202} &=F^{022}_{022}=F^{220}_{220}=F^{022}_{200}=F^{202}_{020}\nonumber\\
&               &=F^{220}_{002}=F^{000}_{222}=F^{013}_{013}=F^{301}_{101}\nonumber\\
&               &=F^{031}_{013}=F^{103}_{303}=F^{031}_{100}=F^{103}_{030}\nonumber\\
&               &=F^{310}_{001}=F^{000}_{313}=F^{301}_{010}=F^{130}_{003}\nonumber\\
&               &=F^{013}_{300}=F^{000}_{131}=F^{103}_{121}=F^{031}_{231}\\
&               &=F^{013}_{213}=F^{301}_{323}=F^{323}_{301}=F^{233}_{033}\nonumber\\
&               &=F^{211}_{011}=F^{121}_{103}=F^{112}_{203}=F^{121}_{012}\nonumber\\
&               &=F^{202}_{111}=F^{022}_{331}=F^{332}_{023}=F^{233}_{302}\nonumber\\
&               &=F^{103}_{212}=F^{031}_{322}=F^{323}_{032}=F^{332}_{201}=F^{202}_{333}\nonumber\\
&               &=F^{112}_{021}=F^{211}_{102}=F^{013}_{122}=F^{301}_{232}=F^{000}_{000}=1.\nonumber
\end{eqnarray}
$F$--symbols admitting two solutions $\pm1$ are as follows.
\bse\label{eq:z4FsymbWithSign}
\begin{eqnarray}
&F^{112}_{112}& =F^{332}_{332}=\pm 1 \\
&F^{130}_{130}& =F^{310}_{310}=d_1^{-1}=\pm 1 \\
&F^{121}_{321}& =F^{211}_{233}=F^{233}_{211}=F^{323}_{123}=\pm d_1=\pm 1\qquad \\
&F^{211}_{320}& =F^{220}_{133}=F^{323}_{210}=F^{310}_{223}=F^{233}_{120}\nonumber\\
& &=F^{220}_{311}=F^{121}_{230}=F^{130}_{221}=v_2^{-1}=\pm 1\\
&F^{310}_{132}& =F^{130}_{312}=F^{332}_{110}=F^{112}_{330}=\frac{v_2}{d_1}=\pm 1.
\end{eqnarray}
\ese
All of the $F$--symbols except the two in the first row in Eq. (\ref{eq:z4FsymbWithSign}) depend on the choice of the sign in either $d_1$ or $v_2$, or both. Note that the $F$--symbols in Eq. (\ref{eq:z4FsymbWithSign}e) are determined by the choice of those in Eqs. (\ref{eq:z4FsymbWithSign}b) and (\ref{eq:z4FsymbWithSign}d). The choice that $F\equiv1$  gives the $F$--symbols of the group $\Z_4$.
\subsection{$\F_{\Z_6}$ $6j$ symbols}\label{appx:6jZ6}
We shall forgo the search of $\F_{\Z_5}$ $6j$-symbols but record the $\F_{\Z_6}$ $6j$ symbols in this section, as $\F_{\Z_6}$ is the simplest case where the three edges incident at a honeycomb vertex can have three different string types that are nonzero and not dual to each other. As proven before, this is the first case where complex $F$--symbols may arise.

There are six string types, including type zero. The branching rules are $\{0,0,0\}$, $\{0,1,5\}$, $\{0,2.4\}$, $\{0,3,3\}$, $\{1,1,4\}$, $\{1,2,3\}$, $\{2,2,2\}$, $\{2,5,5\}$, $\{3,4,5\}$, $\{4,4,4\}$. Clearly, only $0$ and $3$ are self-conjugate.

There are two independent sets of real $6j$ symbols where none of the index takes value zero, each set
generated by the tetrahedral symmetry.

The first set have two independent choices of signs that cannot be constrained by hermiticity of $F$, or any $2G$ or the
full pentagon relations.
\begin{eqnarray}\label{eq:z6Gsymbreal1}
& G^{435}_{411} &=G^{411}_{435}=G^{141}_{523}=G^{525}_{143}=G^{552}_{352}\nonumber\\
&               &=G^{354}_{554}=G^{114}_{314}=G^{312}_{112}=G^{231}_{255}\nonumber\\
&               &=G^{255}_{231}=G^{231}_{255}=G^{123}_{541}=G^{321}_{521}=G^{525}_{325}\nonumber\\
&               &=G^{552}_{534}=G^{534}_{552}=G^{453}_{211}=G^{132}_{114}=G^{114}_{132}\nonumber\\
&               &=G^{411}_{253}=G^{255}_{413}=G^{345}_{145}=G^{213}_{455}=G^{141}_{341}= \pm 1.\nonumber
\end{eqnarray}

The second independent set is given by
\begin{eqnarray}\label{eq:z6Gsymbreal2}
& G^{453}_{453} &=G^{345}_{321}=G^{321}_{345}=G^{132}_{532}=G^{534}_{134}\nonumber\\
&               &=G^{213}_{213}=G^{435}_{235}=G^{231}_{431}=G^{123}_{123}\nonumber\\
&               &=G^{312}_{354}=G^{354}_{312}=G^{543}_{543} = d_1.
\end{eqnarray}
Note that the sign of this set is tied to that of $d_1$ from the full pentagon relations.

A second fact to note is that the indices of each column is either both even, or both
odd. Therefore the resulting $F$ symbols are real. However, if $d_1 = d_3 = -1$,
the resulting $F$ symbols can take either signs, depending on the indices on the last
column, and the relative sign of $v_1$ and $v_3$.

Finally, we have two sets of complex $G$'s that cannot be related by tetrahedral symmetry,
but whose values are related via pentagon relations, or equivalently the
hermiticity of the $F$ symbols.

The first set is

\begin{eqnarray}\label{eq:z6Gsymbcx1}
& G^{132}_{441} &=G^{444}_{135}=G^{444}_{553}=G^{552}_{443}=G^{255}_{322}\nonumber\\
&               &=G^{321}_{254}=G^{312}_{445}=G^{444}_{311}=G^{231}_{522}\nonumber\\
&               &=G^{525}_{234}=G^{123}_{214}=G^{213}_{122}= e^{\ii\alpha_1}.\nonumber
\end{eqnarray}

The second set is
 \begin{eqnarray}\label{eq:z6Gsymbcx1}
& G^{141}_{432} &=G^{435}_{144}=G^{543}_{452}=G^{453}_{544}=G^{345}_{412}\nonumber\\
&               &=G^{411}_{344}=G^{114}_{223}=G^{222}_{113}=G^{354}_{221}\nonumber\\
&               &=G^{222}_{355}=G^{222}_{531}=G^{534}_{225}= e^{\ii\alpha_2}.\nonumber
\end{eqnarray}

The constraint is given by
\be \label{eq:z6Gsymconst}
e^{\ii (\alpha_1+ \alpha_2)} = \frac{1}{d_1} = d_1.
\ee

This means that the two sets are complex conjugates of each other only
if we take $d_1=+1$.
Note that since we have only two even elements which
are related by conjugation at
$N=6$ i.e. 2 and 4,
the complex $G$'s listed above, whose indices in each
column are not equal nor conjugates of each other
must be such that one is even, and the other odd.
This is what led to the weaker constraint in Eq. (\ref{eq:z6Gsymconst}),
rather than the stronger one where the rhs is strictly equal
to plus one that would generally follow from Eq. (\ref{eq:znGs=0}).

A second fact is to notice that one can further separate the $G$
symbols into two classes within each of the real and complex classes described previously.

For class (A), any given column has
both indices taking even (or odd) values. In which case, the indices in each of \emph{all} the
rest of the columns would also either be both even or both odd.

For class (B), one index is even and the other is odd in a given column. That however
implies the same is true for the rest of all the columns. These are simple results
following from the branching rules.

These two different kinds of $G$ symbols cannot be related by tetrahedral symmetry.
Therefore the product $d_{l\s{A}}d_s$ appearing in Eq. (\ref{eq:znGs=0})
can only be $+1$ in class (A), and $-1$ in class (B).

It further implies that complex $G$'s falling into class (B) are such that each component and its Hermitian conjugate are related by tetrahedral symmetry.  These kind of $G$'s only appear beginning at $N=8$.

\subsection{Gauge fixing}

In the previous section where the $6j$ symbols corresponding to $\F_{\Z_6}$ fusion rules
were studied, we have seen that there is a continuous phase parameter
$\alpha$ that cannot be determined from the pentagon relations.
One observation here however, is that suppose one consider further rescalings
where $f_{ijk}\to \tilde{f}_{ijk} = f_{ijk} e^{\ii\theta_{ijk}}\,,$ and similarly
$f_{i*j*k*}\to\tilde{f}_{i*j*k*}= f_{i*j*k*} e^{-\ii\theta_{ijk}}$, the two groups of real $G$ symbols,
with representatives $G^{435}_{411}$ and $G^{453}_{453}$ respectively,
are invariant. This is true provided of course that $f_{ijk}$ is symmetric in all the indices
as is already assumed in our solution of the $6j$ symbol satisfying
reflection symmetry. We note also that the residual rescaling transformation on $G$ is identical
to that of $F$ given in Eq. (\ref{eq:phase1}).

On the other hand, when we inspect the rescaling of the two sets of complex $G$'s, represented
respectively by $G^{132}_{441}$ and $G^{141}_{432}$ we can see that they transform
under rescaling as
\be
G^{132}_{441} \to \frac{e^{\ii \theta_{141}}e^{\ii \theta_{534}}}{e^{\ii \theta_{132}}e^{\ii \theta_{444}}} G^{132}_{441}\,,
 G^{141}_{432} \to \frac{e^{\ii \theta_{123}}e^{\ii \theta_{444}}}{e^{\ii \theta_{141}}e^{\ii \theta_{534}}} G^{141}_{432}.
\ee
Clearly the two groups are rescaled by factors which are precisely complex
conjugates of each other.
As a result, by setting
\be\label{eq:gauge2}
\frac{e^{\ii \theta_{141}}e^{\ii \theta_{534}}}{e^{\ii \theta_{132}}e^{\ii \theta_{444}}} = e^{-\ii\alpha_1},
\ee
we automatically have $G^{132}_{441} = 1$, and $G^{141}_{432}= d_1$.

There are further residual rescaling symmetry by choosing independent phases for $\theta_{141},\theta_{234}$ and $\theta_{444}$,
since the extra gauge choice Eq. (\ref{eq:gauge2}) constrains only their product. However these
redundancies do not lead to further redundancy in physical variables.

In fact, discrete sets of $F$--symbols can also be equivalent up to rescaling, e.g., the $+1$ and $-1$ $\F_{\Z_3}$ $F$--symbols are rescaling equivalent. But we shall offer definite answer to the question which $\F_{\Z_N}$ $F$--symbols belong to the same equivalence classes elsewhere.

\bibliographystyle{apsrev}
% Produces the bibliography via BibTeX.
\bibliography{TensorNets}

\end{document}